\documentstyle{mn}
\input{epsf}
\title[Formation and evolution of structure]
{Statistical characteristics of formation and evolution\\
of structure in the universe.}

\author[Demia\'nski ~\&~ Doroshkevich]
       {M. Demia\'nski$^{1,2}$~\&~ A.G. Doroshkevich$^{3,4}$\\
	$1$Institute of Theoretical Physics,
                       University of Warsaw,
                       00-681 Warsaw, Poland\\
        $2$Department of Astronomy, Williams College,
           Williamstown, MA 01267, USA\\
	$3$Theoretical Astrophysics Center,
          Juliane Maries Vej 30,
          DK-2100 Copenhagen \O, Denmark\\
	$4$Keldysh Institute of Applied Mathematics,
                        Russian Academy of Sciences,
                        125047 Moscow,  Russia}
\date{Accepted ...,
      Received 1998 October ...;
	in original form 1998 October 13}
\begin{document}
\maketitle

\begin{abstract}
An approximate statistical description of the formation
and evolution of structure of the universe based on the Zel'dovich
theory of gravitational instability is proposed. It is found that
the evolution of DM structure shows features of self-similarity and
the main structure characteristics can be expressed through the
parameters of initial power spectrum and cosmological model. For
the CDM-like power spectrum and suitable parameters of the cosmological
model the effective matter compression reaches the observed scales
$R_{wall}\sim $20 -- 25$h^{-1}$Mpc with the typical mean separation
of wall-like elements $D_{SLSS}\sim $50 -- 70$h^{-1}$Mpc. This
description can be directly applied to the deep pencil beam galactic
surveys and absorption spectra of quasars. For larger 3D catalogs
and simulations it can be applied to results obtained with the
core-sampling analysis.

It is shown that the interaction of large and small scale
perturbations modulates the creation rate of early Zel'dovich
pancakes and generates bias on the SLSS scale.
For suitable parameters of the cosmological model and reheating
process this bias can essentially improve the characteristics of
simulated structure of the universe.

The  models with $0.3\leq \Omega_m \leq 0.5$ give the best description
of the observed structure parameters. The influence of low mass "warm"
dark matter particles, such as a massive neutrino, will extend the
acceptable range of $\Omega_m$ and $h$.
\end{abstract}


\section{Introduction}

Over the past decade the large maps of the spatial galaxy distribution
have been prepared and the unexpectedly complicated character of
this distribution was established. The structure predicted by the
Zel'dovich theory of gravitational instability (Zel'dovich 1970, 1978)
was found already in the first wedge diagrams (Gregory ~\&~ Thompson
1978) and now the Large Scale Structure (LSS) is seen in many
observational catalogs, such as the CfA (de Lapparent, Geller ~\&
Huchra 1987; Ramella, Geller \&~Huchra 1992),  the SRSS
(da Costa et al. 1988) and in the Las Campanas
Redshift Survey (Shectman et al. 1996, hereafter LCRS).  The
observed high concentration of galaxies within the wall-like structure
elements such as the Great Attractor (Dressler et al.  1987), and the
Great Wall (de Lapparent, Geller \&~ Huchra 1988) and the existence of
extended under dense regions similar to the Great Void (Kirshner et
al. 1983) put in the forefront the investigation of the
Super Large Scale Structure (SLSS).  Now the
SLSS is also found in many deep pencil beam redshift surveys
(Broadhurst et al. 1990; Willmer et al. 1994; Buryak et al. 1994;
Bellanger \&~ de Lapparent 1995; Cohen et al. 1996) as a rich galaxy
clumps with the typical separations in the range of (60 --
120)$h^{-1}$Mpc. Here h=$H_0$/100 km/s/Mpc is the dimensionless Hubble
constant.

Further progress in the statistical description of the LSS \&~ SLSS
has been reached with the core-sampling method (Buryak et al. 1994)
and the Minimal Spanning Tree technique (Barrow, Bhavsar \&~ Sonoda
1985). Recent analysis of the LCRS performed by Doroshkevich et al.
(1996 \&~1997b, hereafter LCRS1 \&~ LCRS2, Doroshkevich et al. 1998) 
revealed some statistical
parameters of the wall-like SLSS component such as their typical
separation, $D_{SLSS}\approx 50 - 60 h^{-1}$Mpc, and the fraction of
galaxies accumulated by the SLSS, which can reach $\sim$ 50\%. The
same analysis indicates that formation of richer walls can be roughly
described as an asymmetric 2D collapse of regions with a typical size
$R_{wall}\sim$ 20 -- 25$h^{-1}$Mpc that is about half of their typical
separation. The analysis of Durham/UKST redshift survey confirms
these results (Doroshkevich et al., 1999). Earlier
similar scales, in the range of (50 -- 100)$h^{-1}$Mpc, were found
only for spatial distribution of clusters of galaxies (see, e.g.,
Bahcall 1988; Einasto et al. 1994) and for a few superclusters of
galaxies (see, e.g., Oort 1983 a,b).

Evolution of  structure was discussed and simulated many times
(see, e.g., Sahni et al. 1994; Doroshkevich et al. 1997a,
hereafter DFGMM; for references, Sahni \& Coles 1995). However,
SLSS in the dark matter (DM) distribution similar to that seen in
the LCRS was found only recently in a few simulations with the
CDM-like power spectrum and $\Omega_mh$= 0.2 -- 0.3, (Cole, Weinberg,
Frenk \& Ratra 1997; Doroshkevich, M\"uller, Retzalf \& Turchaninov
1999, hereafter DMRT). Hence, for suitable cosmological models the
evolution of small initial perturbations results in the SLSS formation.

In this paper we present an approximate statistical description of
the process of DM structure formation based on the nonlinear Zel'dovich
theory. The potential of this approach is limited as the successive
consideration of mutual interactions of the small and large scale
perturbations becomes more and more cumbersome. In spite of this
it allows us to obtain some interesting results. Thus, it is shown
that formation of both LSS and SLSS is a joint process possessing
some features of self - similarity. The main observed characteristics
of LSS and SLSS are expressed through the structure functions
of power spectrum and through the typical scales, set by the power
spectrum, the time scale, set by the amplitude of perturbation, and
the main parameters of cosmological model. One of the most interesting
such characteristics is the dynamical scale of the nonlinearity
defined as the scale of essential DM concentration within high density
walls. We show that for the CDM transfer function (Bardeen et al. 1986,
hereafter BBKS) and Harrison -- Zel'dovich primordial power spectrum
and for cosmological models with lower matter density this scale of
nonlinearity reaches 20 -- 30$h^{-1}$Mpc that is comparable with typical
scales of the observed SLSS elements.

Simulations (DMRT) show that even in cosmological models with a low
matter density the simulated velocity dispersion within the SLSS
elements reaches 400 -- 700km/s {\it along each principal axis};
this exceeds the observed value by a factor of $\sim$ 1.5 -- 2.
Such a large and isotropic velocity dispersion is caused by the
disruption of the walls into high density clouds. For smaller matter
density of the universe this dispersion decreases but together with
the fraction of matter accumulated by the walls. This means that other
factors as, for example, the large scale bias in the spatial galaxy
distribution relative to the more homogeneous distribution of DM and
baryons could be essential for the successful reproduction of the
observed SLSS. Such  large scale bias caused by the interaction
of small and large scale perturbations was discussed by  Dekel \&
Silk (1986), and Dekel \& Rees (1987), and estimated by
Demia\'nski \&~ Doroshkevich (1999, hereafter Paper I).

The interaction of small and large scale perturbations is
important during all evolutionary stages.
Thus, even during early evolutionary periods the large
scale perturbations modulate the rate of pancake formation. This
modulation is seen as an acceleration of the pancake formation
within deeper potential wells which later are transformed into the
wall-like SLSS elements (Buryak et al. 1992; Demia\'nski \&
Doroshkevich 1997; Paper I). Suppression of  pancake
formation near the peaks of gravitational potential noted by Sahni et al. (1994)
is another manifestation of such interactions. During all evolutionary
stages these interactions result in the successive merging of individual
pancakes. The acceleration of
pancake disruption, caused by  compression of matter within walls,
can also be attributed to this interaction. Now it is observed as
a high velocity dispersion in simulated SLSS and as differences
between the expected and measured mass functions. It was found to
be essential even for pancakes  formed at high redshifts
(Miralda-Escude et al. 1996).
The possible correlation of galaxy morphology with large scale
perturbations was discussed by Evrard et al. (1990). All these
manifestations  of small and large scale interaction
are important for the correct comparison and
interpretation of simulated and observed matter distribution.

Now the modulation of spatial distribution of pancakes
 formed at high redshifts $z\geq 4$
can be seen as the large scale bias in the galaxy and DM
spatial distribution. This bias can be generated by the combined
action of large scale perturbations and reheating of baryonic
component of the universe (see, e.g., Dekel \& Silk 1986; Dekel
\& Rees 1987). The reheating was discussed many times during the
last thirty years in various aspects (see, e.g., Sunyaev \&
Zel'dovich 1972; White \& Rees 1978; Shapiro, Giroux \& Babul
1994). Effects of reheating on the process of galaxy formation
were discussed as well (see, e.g., Babul \& White 1991; Efstathiou
1992; Quinn, Katz \& Efstathiou 1996). It is also known that under
reasonable assumptions about the possible energy sources reheating
can occur for relatively small range of redshifts $z\approx 5-10$
(see, e.g., Tegmark et al. 1997; Baltz et al. 1998).  If
essential concentration of baryons in high density clouds is reached
at the same redshifts, the reheating can help to generate bias
(Demia\'nski \& Doroshkevich 1997; Paper I).
In this case  further formation of high density baryonic clouds
will be significantly depressed, due to reheating, within extended
regions observed today as  under dense regions between richer
walls. Our estimates show that this spatial modulation of the
luminous matter distribution may be essential for the interpretation
of observations.

Numerical simulations are now the best way to reproduce and to study
the joint action of all the pertinent factors together and to obtain
more representative description of the process of structure
formation. Essential progress achieved recently both in the
simulations and study of DM and 'galaxy' distributions (Governato et
al. 1998; Jenkins et al. 1998; Doroshkevich, Fong \&~ Makarova 1998;
DMRT; Cole, Hatton, Weinberg \& Frenk 1998) allows us to follow the
structure evolution in a wide range of redshifts and to reveal
differences between DM and galaxy distribution. Comparison
of these results with observations and an approximate theoretical
description stimulates further progress in our understanding of
evolution of the universe.

This paper is organized as follows: In Sect. 2 main notations are
introduced. In Sec. 3 and 4 the distribution functions of  DM
pancakes are derived and the interaction of small and large scale
perturbations is described that allows us to obtain in Secs. 5 and 6
the statistical characteristics of DM structure. In Sec. 7 the large
scale bias is discussed and in Sec. 8 the dynamical characteristics of
walls are found.  In Sec. 9 the theoretical  estimates are compared
with the available observational and simulated data.  We conclude with
Sec. 10 where a short discussion of the main results is presented. Some
technical details are given in Appendixes I -- IV.

\section{ Statistical parameters of perturbations: variances and
typical scales}

The simplest characteristics of perturbations are the variances of
density and velocity perturbations. For a more detailed statistical
description of the structure evolution it is necessary to use also
the structure functions.  They were introduced in Paper I and are
briefly described in this Section and Appendix I. Here we consider
only the SCDM-like power spectrum but the same approach can be
applied for other spectra as well.

Our analysis is based on the Zel'dovich theory which links the
Eulerian, $r_i$, and the Lagrangian, $q_i$, coordinates of fluid
elements (particles) by the expression
$$r_i = (1+z)^{-1}[q_i - B(z)S_i({\bf q})],\eqno(2.1)$$
where $z$  denotes  the redshift, $B(z)$ describes growth of
perturbations in the linear theory, and the potential
vector $S_i({\bf q})=\partial \phi/\partial q_{i}$ characterizes the
spatial distribution of perturbations. The Lagrangian coordinates of a
particle, $q_i$, are its unperturbed comoving coordinates.

For the flat universe with $\Omega_m+\Omega_\Lambda = 1,~~\Omega_m
\geq 0.1$, the function $B(z)$ can be approximated with a precision
better then 10\% by the expression (Paper I)
$$B^{-3}(z)\approx {1-\Omega_m+2.2\Omega_m(1+z)^3\over
1+1.2\Omega_m},\eqno(2.2)$$
and for an open universe with $0.1\leq \Omega_m\leq 1,~ \Omega_
\Lambda=0$ as
$$B^{-1}(z)\approx 1+{2.5\Omega_m\over 1+1.5\Omega_m}z\eqno(2.3)$$
(Zel'dovich \& Novikov 1983). For
$\Omega_m=1,~~\Omega_\Lambda=0$ both expressions give $B^{-1}(z)=1+z$.

The main characteristics of the perturbations are the
variances of density $\sigma^2_\rho$, displacement $\sigma_s^2$, and
components of the deformation tensor $\sigma_D^2 $
$$\sigma_s^2 ={1\over {2\pi^{2}}}\int_0^\infty p(k)dk,\quad
\sigma^2_\rho =5\sigma_D^2={1\over {2\pi^{2}}}\int_0^\infty p(k)k^2dk,
\eqno(2.4)$$
where $p(k)$ is the power spectrum, and $k$ is the comoving wave number.

The power spectrum determines also two amplitude independent typical
scales, which allow us to describe the process of structure formation 
and can be, possibly, estimated from the observed galaxy distribution.
For the Harrison -- Zel'dovich primordial power spectrum these
scales, $l_0$ and $l_c$, are defined as
$$l_0^{-2}=\int_0^\infty kT^2(k/k_0)dk,\quad
l^2_c={5\over 3}{\sigma_s^2\over\sigma^2_\rho}, \eqno(2.5)$$
where $T^{2}(x)$ is the transfer function and  $k_0=\Omega_mh^2$Mpc
$^{-1}$. For the CDM 
transfer function (BBKS) the scale $l_0$ and the typical masses of 
DM and baryonic components associated with the scale $l_0$ are
$$l_0\approx 6.6(\Omega_mh)^{-1}\sqrt{0.023\over m_{-2}} h^{-1}{\rm
Mpc},\eqno(2.6)$$
$$M_0 = {\pi\over 6}<\rho> l_0^3\approx {2.5\cdot 10^{13}
M_\odot\over \Omega_m^2h^4},\quad M_b^{(0)} = {\Omega_b\over
\Omega_m}M_0.$$
$$m_{-2} = \int_0^\infty xT^2(x)dx$$

For the SCDM power spectrum the $\sigma^2_\rho$ is divergent
(logarithmically) at large $k$ and the standard CDM transfer function
should be truncated by introduction of an appropriate mass of DM
particle, $M_{DM}$. This restriction allows, however, a significant
interval for
a possible mass of DM particles -- up to $10^{16}$ -- $10^{20}$~eV.
In this paper we will use the ratio $l_c/l_0$ as a measure of
the mass of DM particles. For 0.5~keV $\leq M_{DM}\leq 10^{17}$~keV
we have 0.2 $\geq l_c/l_0\geq 0.01$  and for $M_{DM}\geq 5$keV the
following approximate expression can be used
$$l_c/l_0\approx {0.061\over\sqrt{\ln M_{DM}}},\eqno(2.7)$$
$$M_c = \left({l_c\over l_0}\right)^3M_0\approx
{0.6\cdot 10^{10}M_\odot\over (\ln M_{DM})^{3/2}\Omega_m^2h^4},$$
where $M_{DM}$ is the mass of DM particle in keV.
It turns out that the final results
only weakly depend on the mass of DM particles. For the
Harrison-Zel'dovich primordial power spectrum the dependence of the
moments and typical scales on the mass of DM particles was discussed
in Paper I. More cumbersome correlation and structure functions of
perturbations were also discussed in Paper I and, partly, are
presented in the Appendix I.

The amplitude of power spectrum can be taken from the measured
anisotropy of the relic radiation (Stompor et al. 1995; Bunn~\&~White
1997; Gorski et al. 1998). It can be fitted by
$$\sigma_s\approx 31.2\sqrt{m_{-2}\over 0.023}
\left({T_Q\over 20\mu K}\right)a(\Omega_m,\Omega_\Lambda){\rm Mpc},\eqno(2.8)$$
$$a(\Omega_m,\Omega_{\Lambda}=1-\Omega_m) =
\Omega_m^{0.215-0.05ln\Omega_m},$$
$$a(\Omega_m,\Omega_{\Lambda}=0) = \Omega_m^{0.65-0.19ln\Omega_m}.$$
where $T_Q$ is 
the amplitude of quadrupole component of anisotropy of the relic
radiation. The time scale of the structure evolution is defined by the
function
$$\tau(z) = \tau_0B(z),\eqno(2.9)$$
$$ \tau_0={\sigma_{s}\over\sqrt{3}l_0}\approx
2.73h^2 \Omega_m \left({m_{-2}\over 0.023}
{T_Q\over 20\mu K}\right)a(\Omega_m,\Omega_\Lambda),$$
$$\tau_0\approx 2.73h^2\Omega_m^{1.21}\left({m_{-2}\over 0.023}
{T_Q\over 20\mu K}\right),
~~\Omega_\Lambda=1-\Omega_m,$$
$$\tau_0\approx 2.73h^2\Omega_m^{1.65-0.19ln\Omega_m}\left({m_{-2}\over 0.023}
{T_Q\over 20\mu K}\right),
~ \Omega_\Lambda=0, ~\Omega_m\leq 1.$$

Further on, as a rule, the dimensionless variables will be used.
We will use $l_0$ as the unit of length, and $<\rho> l_0 $
as the unit of surface density. This means also that such dimensionless
characteristics of a pancake as the size of collapsed slab and
resulting surface density of a pancake are identical. Below we will use
both terms as well as the term 'mass', $m$, to characterize the
surface density reached during formation of a pancake. The gravitational
potential and displacement are measured in units of
$\sigma_s l_0/\sqrt{3}$ and $\sigma_s/\sqrt{3}$, respectively.

\section{Statistical characteristics of pancakes}

In this section the mass function of  Zel'dovich pancakes
and its time evolution is given. This can be done using the
main equation of Zel'dovich theory (2.1). The mass of compressed
matter is measured by the Lagrangian size of compressed slab, $q$,
or by the surface density of pancake, $<\rho>q$. As it was noted
above, both measures are identical in dimensionless notation. So,
we will use both terms 'size', $q$, and 'mass', $m$, of a pancake
to characterize its surface density.

Here we do not consider the transversal characteristics of structure
elements and cannot discriminate, for example, the central part of
a poorer pancake and periphery of richer pancake, if they have the
same surface density or mass, $m$. In this sense our approach gives
characteristics similar to that obtained with the core-sampling
analysis, pencil beam observations or the distribution
of absorption lines in spectra of QSOs.

\subsection{The pancake formation}

According to the relation (2.1) when two particles with different
Lagrangian coordinates $\bf q_1$ and  $\bf q_2$ meet at  the same
Eulerian point $\bf r$ a pancake with the surface mass density
$<\rho>|{\bf q}_1-{\bf q}_2|$ forms. Here we assume that
{\it all} particles situated between these two boundary
particles are also incorporated into the same pancake. This
assumption is also made in the adhesion approach (see, e.g., Shandarin
\& Zel'dovich 1989).
Formally, this condition can be  written as
$${\bf q}_{12} = {\bf q}_1-{\bf q}_2 = \tau\cdot[{\bf S}({\bf q}_1)-
{\bf S}({\bf q}_2)].\eqno(3.1)$$
This means that the pancake formation
process can be characterized by the scalar random function
$$Q(q_{12}) = {{\bf q}_{12}\over q_{12}}\cdot[{\bf S}({\bf q}_1)-
{\bf S}({\bf q}_2)]\eqno(3.2a)$$
under the condition
$${\bf q}_{12}\times [{\bf S}({\bf q}_1)-
{\bf S}({\bf q}_2)] = 0.$$
In a coordinate system with the first axis oriented along the vector
${\bf q}_{12}$ the condition (3.1) can be rewritten more explicitly as
$$Q(q_{12}) = \Delta S_1= q_{12}/\tau, ~~ \Delta S_2(q_{12}) =
\Delta S_3(q_{12}) = 0,\eqno(3.2b)$$
$$\Delta S_i=S_i({\bf q}_1)-S_i({\bf q}_2),$$
and  formation of a pancake is described by the 1D
matter flow between points $q_1 = {\bf q}_1\cdot{\bf q}_{12}/q_{12}$
and  $q_2 = {\bf q}_2\cdot{\bf q}_{12}/q_{12}.$

As is shown in Appendix II, under the assumption of Gaussian
distribution of perturbations and neglecting the (weak) correlations
between matter motion in orthogonal directions,
the probability of a pancake formation, for a given $q$
prior to the 'time' $\tau$ or for a given $\tau$ with a size
larger then $q$, is identical to the probability to have
$\Delta S_1\geq q/\tau$ ($Q(q_{12})/q_{12}\geq 1/\tau$), so
$$W_{cr}(>q,\tau) = 1-{1\over 8}\left[1+erf\left({\mu(q)\over
\sqrt{2}\tau}\right)\right]^3,\eqno(3.3)$$
$$\mu(q)={q\over \sqrt{2[1-G_{12}(q)]}}.$$
where (see Appendix I \& II)
$$\mu(q)\approx\sqrt{q_0/3}~,~~q< q_0,$$
$$\mu(q)\approx \sqrt{q}/2,~~q_0\ll q< 1,\quad
\mu(q)\approx q/\sqrt{2},~~q\gg 1,\eqno(3.4)$$
$$q_0\approx 6l_c^2/l_0^2\approx {0.022\over\ln M_{DM}},\quad
M_q = q_0^3M_0\approx {3.\cdot 10^{8}
M_\odot\over (\ln M_{DM})^3\Omega_m^2h^4}.$$
As is seen from (3.1), and (3.3) the pancake characteristics at
the moment $\tau$ are expressed through the function $\mu(q)$
related to the initial power spectrum. This is an essential
feature of the Zel'dovich theory allowing one to obtain approximate
analytical description of pancake properties.

Equation (3.3) demonstrates that $\sim{7\over 8}$ of matter with
$Q(q_{12})\geq 0$ is compressed at least in one direction and for
$\sim{1\over 8}$ of matter 3D expansion takes place what reflects
the symmetry of initial distribution of the displacements,
$\Delta S_i$. At first glance it seems that this conclusion is in
contradiction with known results of Zel'dovich theory. Actually,
the fraction of mass, $f_{DM}$, accumulated by pancakes can be
estimated using the PDF for the largest principal value
of deformation tensor, $dW(\lambda_1)$, (Zel'dovich 1970; Doroshkevich
1970; Doroshkevich \& Shandarin 1979; Shandarin \& Zel'dovich 1989;
Paper I). This PDF
can be approximated (with a precision $\sim$10\%) by the Gaussian
function with $<\lambda_1>=3\sigma_D/\sqrt{2\pi},~~
\sigma_\lambda^2 = (13/6-4.5/\pi)\sigma_D^2$.
As in the Zel'dovich theory pancake formation is described by the
relation $B(z)\lambda_1= 1$ (which follows directly from (2.1)~),
so for the fraction of compressed matter with $\lambda_1\geq 1/B$ 
we have
$$f_{DM}\approx 
{1\over2}erfc\left({\sigma_D\over\sqrt{2}\sigma_\lambda}
\left({l_c\over l_0\tau} - {3\over\sqrt{2\pi}}\right)\right)
\eqno(3.5)$$
and for $\tau\gg l_c/l_0$, $ f_{DM}\rightarrow$1 (in the Zel'dovich theory
 0.92$\leq f_{DM}\leq$1). This shows that already during
the early  period of nonlinear evolution, at $\tau\approx l_c/l_0\ll$ 1,
large fraction of matter $f_{DM}\geq$ 0.9 is compressed
into low mass pancakes. But for the CDM-like power spectrum the
description of matter compression through the deformation tensor
is appropriate only at small scales $q\leq q_0$ whereas for
$q\gg q_0$, the correlations between matter flow in orthogonal
directions rapidly decrease
what can be seen directly from the expressions for the structure
functions given in Appendix I.

At larger scales we have to use the
more cumbersome description discussed above and the estimate (3.3)
shows that at such scales the efficiency of matter integration into
structure elements is only $\sim$ 0.875 (more accurate estimates
taking into account the correlation of displacements lowers this
value to  0.79). This limit is reached already
at small $\tau$, for $q\ll 1$, that means strong matter
concentration within small structure elements.  Further evolution does
not change this limit and only redistributes -- due to sequential
merging -- the compressed matter to more and more massive structure
elements. Thus, the approximate estimates show that $\sim$ 21\% of
matter is subjected to 3D compression, $\sim$ 21\% to 3D expansion,
 $\sim$ 29\% of
matter is subjected to 2D compression and can be accumulated by
filaments and $\sim$ 29\% is subjected to 1D compression and
remains in pancakes.
The difference between estimates (3.3) and (3.5) shows that
$\sim$ 15 -- 20\% of matter incorporated in small clouds, with
$M\leq M_q$, is not accumulated by larger pancakes,
with $M\geq M_q$, and remains distributed between those pancakes.
These estimates can be changed because the
Zel'dovich approximation becomes invalid when strong
matter compression is reached during pancakes formation.

\subsection{The characteristics of pancake formation}

The probability distribution function (PDF) for pancakes
formed at the moment $\tau$ can be found from (3.3) as
$$N_{cr}(q,\tau) = -{8\over 7}{dW_{cr}\over dq} =
{6\over7\sqrt{2\pi}\tau}{d\mu\over dq}
\Phi\left({\mu\over\sqrt{2}\tau}\right),\quad 0<q<\infty,
\eqno(3.6)$$
$$\Phi(x) =e^{-x^2}[1+erf(x)]^2.$$
Simple analysis based on relations (3.3), (3.6) shows
that the maximum of the PDF (3.6) is reached at $q\approx
1.5q_0\ll$1.

The mean mass (surface density) of formed pancakes is
described by the expression
$$<m(\tau)>= \int_0^\infty qN_{cr}dq = {8\over 7}\int_0^\infty
W_{cr} dq,\eqno(3.7)$$
$$<m(\tau)>\approx 4\tau^2,~~\tau\ll 1,\quad <m(\tau)>\approx
\tau,~~\tau\geq 1,$$
and the mass distribution is characterized by the function
$$N_{cr}^{(m)} = {6\over 7\sqrt{2\pi} \tau }{q\over <m>}
{d\mu\over dq}~\Phi\left({\mu\over\sqrt{2}\tau}\right).
\eqno(3.8)$$

The rate of formation of pancakes with mass $q$ is
$$N_\tau={8\over 7}{dW_{cr}\over d\tau}(q,\tau)={6\over 
7\sqrt{2\pi}}
{\mu\over \tau^2}\Phi\left({\mu\over\sqrt{2}\tau}\right),
\eqno(3.9)$$
and the maximal rate is reached for $\mu(q) \approx \tau.$

These relations take into account the growth of pancakes size
due to accretion of matter and form a reliable basis for
further theoretical considerations of various characteristics of
structure elements. However they cannot be applied directly to
the observed and simulated matter distribution as they do not
consider the possibility of merging of pancakes. In the process
of merging, earlier formed pancakes are accumulated by larger ones.
A more refined technique taking into account the pancake interaction
should be used in order to describe the evolution of {\it structure
elements} that are the pancakes that survived the merging process.
The same problem appears in the Press-Schechter theory and it
can be solved by introduction of the survival probability (Peacock
\& Heavens 1990; Bond et al. 1991). This problem will be discussed
below.

\subsection{Spatial characteristics of pancake distribution}

Using the standard technique (see, e.g., BBKS) we can also find
the mean cumulative comoving linear number density of pancakes,
$n(>q)$, which characterizes the pancake distribution along a random
straight line
$$n(>q) dr = {3\over 4\sqrt{2\pi}}{\mu(q)\over q}<\nabla Q\cdot d\vec{r}>
\Phi\left({\mu\over\sqrt{2}\tau}\right).$$
When $l_c\ll l_0$ it is described by the following (approximate)
expression:
$$n(>q)\approx {3<\mu_r>\over 4\pi \sqrt{q_0}}[\sqrt{6}+
\ln(\sqrt{3}+\sqrt{2}~)]{\mu(q)\over
q}\Phi\left({\mu\over\sqrt{2}\tau}\right),$$
$$l_c n(>q)\approx {1.35\over \sqrt{6} \pi}{\mu(q)\over q}
\Phi\left({\mu\over\sqrt{2}\tau}\right),\eqno(3.10)$$
where $<\mu_r>$= 0.5 describes the random orientation
of a pancake and the chosen straight line (see also  Appendix II).

To characterize the 3D distribution of higher peaks of the function
$Q(q)$ and the percolation process, in the course of which separate
pancakes are integrated into the joint structure, the standard
technique using the Euler characteristics can be applied. It is
described in details by BBKS; Gott et al. (1989); Coles et al. (1996);
Seto et al. (1997).

In our case the anisotropy of the pancake compression manifests
itself as an anisotropy of correlation functions of displacements
and their derivatives what results in the appearance of
different Euler characteristics, $n_{31}(q,\tau), n_{32}(q,\tau),
n_{33}(q,\tau)$:
$$n_{31} = {3\over 32\pi^2}{<Q_{22}Q_{33}-Q_{23}^2>\over
\sigma_{11}\sigma_{d1}\sigma_{d2}\sigma_{d3}}
<|Q_1|>\Phi\left({\mu\over\sqrt{2}\tau}\right),$$
$$Q_i = {\partial Q\over \partial q_i},\quad
Q_{ij} = {\partial^2Q\over \partial q_i\partial q_j}.\eqno(3.11)$$
 where variances $\sigma_{11}, \sigma_{d1}, \sigma_{d2}, \sigma_{d3}$
are defined by
(II.4) and (II.6)  and expressions for $n_{32} = n_{33}$ can
be found by cyclic permutation of indices of derivatives $Q_i \& Q_{ij}$.

Functions $n_{32}~\&~n_{33}$
characterize the percolation process in the direction ${\bf q}_{12}$,
and $n_{31}$ characterizes this process in transversal
directions. For the most interesting case $q_0\ll 1, ~q_0\ll q$
we have
$$l_0^3n_{32} = l_0^3n_{33} = n_0\left({\mu^2\over\tau^2} -
{1+2q\over 3(1+q)}\right),$$
$$l_0^3n_{31} = n_0\left({\mu^2\over\tau^2} - 1\right),\eqno(3.12)$$
$$n_0 = {3\over 4\pi^2}\sqrt{3\over 2}\left({\mu\over qq_0}
\right)^{3/2} \Phi\left({\mu\over\sqrt{2}\tau}\right).$$
The function $n_{31}$ is similar to the standard expression for
an isotropic Gaussian field (BBKS).

The first zero of Euler characteristic describes approximately
the percolation when separate higher peaks are incorporated into
a  larger (in the limiting case - infinite) structure element
(Tomita 1990; Mecke and Wagner 1991). The expressions (3.12) show
that in the directions orthogonal to ${\bf q}_{12}$ the percolation
takes place at the 'moment' $\tau = \mu(q)$ whereas along ${\bf q}_{12}$
it occurs later, at $\tau = \mu(q)\sqrt{3(1+q)/(1+2q)}$. This
fact favors two step percolation and percolation along
the pancake surface takes place first.
For a given $q$ and $\tau$ the ratio $n(>q)/n_{31}$ and $n(>q)/n_{32}$
measure the mean surface of formed pancakes.

Expressions (3.7) \& (3.12) show that at small $\tau$
percolation takes place for pancakes with $m~\approx ~<m>~\approx
~4\tau^2$. This means that only the highest peaks with $m~\geq ~<m>,
~\mu~\geq ~\tau$ can be considered as discrete objects.
The number density of such peaks can be obtained from
(3.12), for $\mu/\tau >$ 1, as
$$l_0^3n_{pk}(>q,\tau) = n_0{\mu^2\over\tau^2}.\eqno(3.13)$$
Later, when objects with $m\sim$ 1 are formed and an essential
fraction of matter is concentrated in
larger pancakes with $m\sim~ <m>~\geq \tau$, the percolation takes place
through relatively low mass pancakes with $m\leq~ <m>$. This ideal
picture will be however strongly distorted by
the matter compression or expansion in
transversal directions and by pancake disruption due to the
gravitational instability.

\subsection{Pancake interaction}

An important characteristic of pancake interaction is the
two points PDF which gives the conditional probability to form a
pancake with the size $q$ at the moment $\tau$ at the distance
${\bf D}_{sep}$ from a pancake with the size $D_1$ formed at
the moment $\tau_1$. Here we will restrict our consideration to
the simplest case when all three vectors, ${\bf q},~~{\bf D}_1, \&
{\bf D}_{sep}$ are aligned along the same straight line that is
the most interesting case for the small separations. Much more
cumbersome case of arbitrary arrangement and/or orientation of
these vectors will be considered separately.

In this simplest case the conditional probability to have
$\Delta S~>q/\tau$ is
$$W_c\left({>q,\tau};{D_1\over \tau_1},D_{sep}\right)  =
0.5\cdot erfc(g_{2}/\sqrt{2}),\eqno(3.14)$$
$$g(x,y,r_s) = {x - r_s y\over \sqrt{1-r_s^2}},\quad
g_2 = g[\mu(q)/\tau,\mu(D_1)/\tau_1,r_s].$$
The function $\mu(q)$ is given by (3.3) and the coefficient of
correlation, $r_s(q,D_1,D_{sep})$, describing the interaction of
two pancakes, can be found in Appendix III (III.1).
The conditional distribution function of such pancakes
can be written as
$$N_c\left({q\over \tau};{D_1\over \tau_1},D_{sep}\right)  =
\sqrt{2\over\pi}{dg_2\over dq}e^{-0.5g^2_2}.\eqno(3.15)$$
These relations characterize the parameters of pancakes formed
at the moment $\tau$ under the condition of a pancake formation at
the moment $\tau_1$ with the size $D_1$ and the separation
$D_{sep}$.

The basic relation (3.1) implies that two pancakes with
sizes $D_1$ and $D_2$ and a separation $|D_{sep}|\leq 0.5(D_1+D_2)$
merge together and form a single pancake. For larger
separations merging of pancakes can also be considered in
the same manner as before, but using the Euler position of formed pancake
$$r_{pan} = \int_{q_2}^{q_1}{r(q)\over q_{12}}dq =
{1\over 1+z}\left[q_{cent} -B(z)
{\Delta\phi_{12}\over q_{12}}\right],\eqno(3.16)$$
where $q_{cent}=0.5(q_1+q_2)$ is the Lagrangian position of
central point of the pancake and 
$\Delta\phi_{12}=\phi(q_1)-\phi(q_2)$. As it is apparent from (3.16),
$r_{pan}$ depends on the potential difference, $[\phi(q_1)-
\phi(q_2)]/q_{12}$, and therefore the pancake evolution depends
on the potential distribution.

As discussed in Paper I around each pancake there is a potential
well with a typical size exceeding the mass $m$ of collapsed
pancake by a factor of 1.5 -- 2. The matter infall into these
potential wells stimulates merging of pancakes formed
with a small separation. For larger separation the influence of
this factor decreases progressively and the process of merging
becomes (almost) random. These results are consistent with predictions
of the adhesion model (Shandarin \& Zel'dovich 1989).

The conditional probability that two pancakes with
sizes $D_1$ and $D_2$ whose separation is $|D_{sep}|\geq 0.5(D_1+D_2)$
merge at the moment $\tau$ is defined by the condition
$$r_{pan}(D_1,\tau,q_{cent}) = r_{pan}(D_2,\tau,q_{cent}+D_{sep}),$$
and instead of eq. (3.3) we have for the probability of merging
$$W_{merg}(D_1,D_2,D_{sep})={1\over 2}erfc\left({\chi(D_1,D_2,
D_{sep})\over\tau\sqrt{2}}\right)\eqno(3.17)$$
The function $\chi(D_1,D_2,D_{sep})$ is given by
(III.6). We can extend applicability of the formula (3.17) for small  
separations by requiring that
$$W_{merg}(D_1,D_2,D_{sep}) =1,\quad
|D_{sep}|\leq 0.5(D_1+D_2).\eqno(3.18)$$

\section{Pancake evolution and formation of filaments}

Similar technique can also be used to estimate the transversal
size, the pancakes compression and/or expansion in transversal
directions, and other properties of pancakes. As we are interested
in the formation of structure elements with typical sizes $q\gg q_0$
the local description through the deformation tensor cannot be used
and the imposed conditions make even approximate description of
pancake evolution quite cumbersome (see, e.g., Kofman et al. 1994).
The general tendencies and rough characteristics of this evolution
can only be outlined. Thus, for example, we can estimate the matter
fraction compressed within filamentary-like elements and high density
clumps as of about 50\% whereas only $\sim$ 29\% of matter is 
subjected to 1D compression. This estimate implies that larger 
pancakes could also incorporate an essential fraction of filaments 
and clumps.

The formation of filaments as well as  pancakes disruption are
stimulated by the growth of density in the course of pancakes
compression, and therefore, probably, during early evolutionary
stages, filaments represent the most conspicuous elements of the
structure. As was discussed in Sec. 3.3, filaments merge to form
a joint network. Later, when larger pancakes are formed, the evolution
of pancakes becomes slower and disruption of pancakes dominates.
These expectations are consistent with the
observed and simulated matter distribution. Thus, the conspicuous
filaments are seen even at $z=3$ (see, e.g., Governato et al. 1998;
Jenkins et al. 1998) whereas disrupted walls dominate at small
redshifts (LCRS1, LCRS2, DMRT).

\subsection{The characteristics of filament formation}

Some approximate characteristics of filament distribution can be
obtained by considering the formation of filaments as a sequential
matter compression along two principal directions. Such two step
compression results in formation of high density "ridge" surrounded
by a lower density anisotropic halo. In a coordinate system with the
first and second axes oriented along the directions of maximal and
intermediate compressions this process can be approximately described
by two equations similar to (3.2):
$$Q(q_{12}) = q_{12}/\tau,\quad Q(y_{12}) = y_{12}/\tau_f.\eqno(4.1)$$
Here vectors ${\bf q}_{12}$ and ${\bf y}_{12}$ and functions
$Q(q_{12})~\&~Q(y_{12})$ describe the deformation along the first
and second coordinate axes respectively and additional conditions
introduced in (3.2) are assumed to be fulfilled.
As was discussed above the matter compression along the second axis
is accelerated by the pancake formation and {\it the function
$\tau_f(z)$ differs from that given by (2.9)}.

Bearing in mind these restrictions we will approximately characterize
the probability of filament formation, $W_{cr}^{f}$, for given
$q~\&~y$ prior to the 'time' $\tau~\&~\tau_f$ or for given
$\tau~\&~\tau_f$ with sizes larger then $q~\&~y$, respectively,
as the probability to have $Q(q_{12})/q_{12}\geq
Q(y_{12})/y_{12}\geq 1/\tau_f,$ $Q(q_{12})/q_{12}\geq 1/\tau:$
$$W_{cr}^{f}(>q,\tau;>y,\tau_f) = 1-{1\over 8}\left[1+erf\left({\mu(q)
\over\sqrt{2}\tau}\right)\right]^3$$
$$-{3\over 8}erfc\left({\mu(q)\over
\sqrt{2}\tau}\right)\left[1+erf\left({\mu(y)\over\sqrt{2}\tau_f}
\right)\right]^2.\eqno(4.2)$$
The PDF for the filaments, $N_{cr}^{f}$, can be found from (4.2) as
$$N_{cr}^{(f)}(q,\tau;y,\tau_f) =
{8\over 3\pi\tau\tau_f}{d\mu(q)\over dq}{d\mu(y)\over dy}
\exp\left[-{\mu(q)\over 2\tau^2}-{\mu(y)\over 2\tau_f^2}\right]\times$$
$$\left[1+erf\left({\mu(y)\over\sqrt{2}\tau_f}\right)\right],
\quad 0<q<\infty,\quad 0<y<\infty\eqno(4.3)$$

More adequate characteristic of filaments is the linear mass density,
$m_f = \pi q y/4$, defined as a mass per unit length of filament.
Both in observed and simulated catalogues filaments usually form
a less massive part of structure elements with $m_f\ll$1. For such
filaments $q\ll 1,$ $\mu(q)\approx \sqrt{q}/2,$ $y\ll 1,$
$\mu(y)\approx \sqrt{y}/2$ and
the PDF $N_{cr}^{f}(m_f)$ can be found from (4.3) as
$$N_{cr}^{f}(m_f,\tau,\tau_f)\approx {1\over 2\pi <m_f>\sqrt{\xi}}
\int_0^\infty {dx\over x}e^{-{\sqrt{\xi}\over 2}(x+1/x)}\eqno(4.4)$$
$$\xi = {m_f\over <m_f>},\quad <m_f> = 4\pi\tau^2\tau_f^2.$$

The surface density of spatial distribution of filaments can be
obtained using the same technique as in Sec. 3.3 .

\subsection{The pancake evolution}

Separation of filaments into a special class of structure elements
requires also a redefinition of population of 'true' pancakes.
Now the probability of 'true' pancake formation defined by
conditions $Q(q_{12})/q_{12}\geq 1/\tau, \quad Q(y_{12})
/y_{12}\leq 1/\tau_f$ is:
$$W_{cr}^{(p)}(>q,\tau;<y,\tau_f) = {3\over 8}erfc\left({
\mu(q)\over\sqrt{2}\tau}\right)\left[1+erf\left({\mu(y)\over
\sqrt{2}\tau_f}\right)\right]^2.$$
instead of expressions (3.2) and (3.3). The PDF for such pancakes is
$$N_{cr}^{(p)}(q,\tau)=\sqrt{2\over\pi}{d\mu(q)\over\tau dq}
\exp\left(-{\mu^2\over 2\tau^2}\right),\quad
0\leq q\leq\infty.\eqno(4.5)$$
$$<m_p> = 4\tau^2\quad q\ll 1.$$
This PDF differs from (3.6) by the form of the function
$\Phi(x)$ what decreases the PDF for larger $\mu/\tau$.

As before, expression (4.5) characterizes pancakes by the surface 
mass density of collapsed matter, $q$. However the pancake surface mass 
density varies with time after pancake formation due to
transversal compression or expansion that results, in particular, in
formation of filaments. Even if these transversal motions do not lead
to such dramatic results they can change drastically the {\it
observed} surface density of pancakes. So, the current surface mass
density $m_p=q/s_p$, where $s_p$ describes the variation of pancake
surface caused by transversal motions, is a more adequate
characteristic of pancakes.

As was discussed above this period of pancake evolution is not
adequately described by the Zel'dovich theory and our results become
unreliable. To obtain qualitative characteristics of
influence of these factors we can, for example, describe the variation
of pancake's surface as
$$s_p\propto (1-\tau^*Q(y_{12})/y_{12})(1-\tau^*Q(z_{12})/z_{12}).$$
Here vectors ${\bf y}_{12}$ and ${\bf z}_{12}$, and functions
$Q(y_{12})~\&~Q(z_{12})$, describe the deformation along the second
and third coordinate axes respectively. The function $\tau^*$ differs
from that given by (2.9) and it can  depend on transversal
motions. Even so rough consideration shows that the exponential term
in (4.5) is eroded, and the resulting PDF becomes power-like:
$$N(m_p\ll 1)\propto m_p^{-1/2},\quad 
N(m_p\gg1)\propto m_p^{-2}.\eqno(4.6)$$
The pancake disruption accelerates this erosion as well and makes
the PDF more complicated.

This discussion shows that slowly evolving pancakes with slower
transversal motions can be separated into a special subpopulation
for which the surface density changes slowly, $m_p\approx q$, and
the PDF (4.5) correctly describes  the pancake distribution during the
essential period of evolution. This subpopulation is singled out by
conditions
$$|Q(z_{12})/z_{12}|\leq |Q(y_{12})/y_{12}|\leq \epsilon/\tau,
\quad \epsilon< 1$$
and the probability of existence of such pancakes is proportional to
$\tau^{-2}$. This factor describes the disappearance of such pancakes
in the course of evolution. This subpopulation can be however quite 
rich (see., discussion in Sec. 9).

\section{Statistical characteristics of dark matter
structure elements}

Results obtained above allow us to find the approximate PDF and
other characteristics of structure elements. The {\it structure
element} with a size (mass) $m$ is defined as a pancake with the
size $m$ formed at a moment $\tau$ that not merged with any
other pancake.

\subsection{Merging of dark matter structure elements}

As before the characteristics of {\it structure
element} at a moment $\tau$ are expressed through characteristics
of initial perturbations. The approximate expression for the PDF
of structure element can be written in a form similar to the known
equation of coagulation:
$$N_{str}(m,\tau) = \eqno(5.1)$$
$$\int_0^\infty dy\int_0^m dx N_{cr}(x,\tau)
N_c(m-x,x,y,\tau){dW_{merg}(x,m-x,y)\over dy}$$
$$-N_{cr}(m,\tau)\int_0^\infty dx \int_0^\infty dy
N_c(x,m,y,\tau){dW_{merg}(x,m,y)\over dy}.$$
The functions $N_{cr}, N_c~\&~W_{merg}$ are given by (3.6), (3.15),
(3.17) \& (3.18). Here the first term describes the formation of two
pancakes with sizes $x~\&~m-x$ and a separation $y$ and their
merging to a pancake with the size $m$ while the second term
describes merging of the pancake of size $m$ with
another pancake. If the mass exchanged during merging
is incorporated into the forming pancake then the first term
in (5.1) has to be appropriately changed.

Here, as the first step of investigation, we will use the
simpler approximate approach based on the survival probability
of a pancake with  size $m$ to avoid merging with larger pancakes
with sizes $x\geq m$. For the more interesting case of smaller pancakes
with $q_0\ll m\leq x<$1, the most probable process is the formation of
two pancakes with sizes $m~\&~x\geq m$ at a small separation
$|D_{sep}|\leq 0.5(x+m)$ that means, as follows from (3.1), formation 
of one structure element with a size $x$. This process is described
by the second term in (5.1). Because in this case
the probability of merging  quickly decreases for larger separations
$|D_{sep}|\geq 0.5(m+x)$ and the function $W_c(m,x,D_{sep})$ weakly depends 
on $D_{sep}$, we will use the approximate expression
for the probability of merging, $P_{mrg}$,
$$P_{mrg}(m,\tau)\approx 2 W_c\left({m, \tau};{m\over \tau},
{m\over \tau}\right),\eqno(5.2)$$
$$P_{mrg}(m,\tau)\approx erfc\left({\mu(m)\over
\tau\sqrt{2}}\right), \quad m\ll 1, \eqno(5.3)$$
and for the survival probability, $P_{surv}$,
$$P_{surv}(m,\tau)\approx 1 - P_{mrg}(m,\tau), \eqno(5.4)$$
$$P_{surv}(m,\tau)\approx  erf\left({\mu(m)\over
\tau\sqrt{2}}\right),\quad m\ll 1.$$
In spite of the approximate character of this approach, it allows
us to obtain reasonable estimates of the expected efficiency of
merging and of the large scale bias. As it is directly seen from
(5.4), for $\mu(m)\leq \tau$,
$$P_{surv}(m,\tau)\propto \mu(m)\tau^{-1},$$
what characterizes the impact of pancake merging.

\subsection{Statistical characteristics of structure elements}

With this survival probability the approximate PDF for the 
structure elements, $N_{str}(q,\tau)$, can be written as follows:
$$N_{str}(m,\tau)\propto  P_{surv}(m,\tau) N_{cr}(m,\tau).
\eqno(5.5)$$
These relations allow us to obtain also the approximate mass
distribution function, $N_{str}^{(m)}(m,\tau)$, characterizing the
distribution of compressed matter over the structure elements.
For the more interesting case $m\ll$1 we have
$$N_{str}(m,\tau)\approx {24\over 17\sqrt{2\pi}\tau}{d\mu\over dm}
\Phi\left({\mu\over \sqrt{2}\tau}\right)\cdot
erf\left({\mu\over\sqrt{2}\tau}\right),\eqno(5.6)$$
$$<m>\approx 8\tau^2.$$
In the general case this function can be obtained numerically.
The mean cumulative comoving linear number density of structure
elements, $n_{str}(>m,\tau)$, is
$$n_{str}(>m,\tau)\approx -\int_m^\infty dm {dn(>m,\tau)\over dm}
P_{surv}(m,\tau)\eqno(5.7)$$
$$= n(>m,\tau)P_{surv}(m,\tau)+\int_m^\infty dm n(>m,\tau)
{dP_{surv}(m,\tau)\over dm}.$$

\begin{figure}
\centering
\epsfxsize=8 cm
\epsfbox{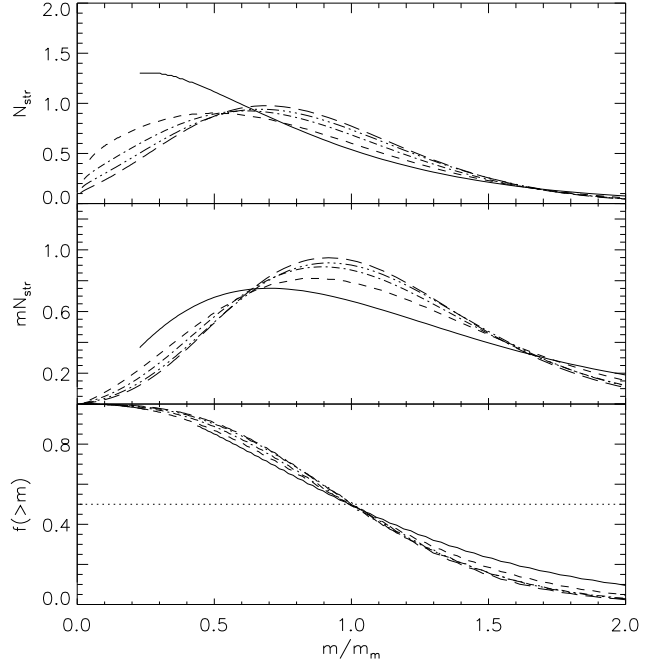}
\vspace{0.75cm}
\caption{The PDF of structure elements, $N_{str}(m,\tau)$,
the mass distribution function, $mN_{str}(m,\tau)$, and the
fraction of compressed matter, $f(>m)$, vs. the masses (sizes)
of structure elements, $m/m_m$,
are plotted for five moments of time: $\tau$ = 0.1 (solid line),
$\tau$ = 0.3 (dashed line), $\tau$ = 0.5 (dot-dashed line),
$\tau$ = 0.7 (dot-dot-dot-dashed line), $\tau$ = 2 (long dashed
line). }
\end{figure}

\begin{figure}
\centering
\epsfxsize=7 cm
\epsfbox{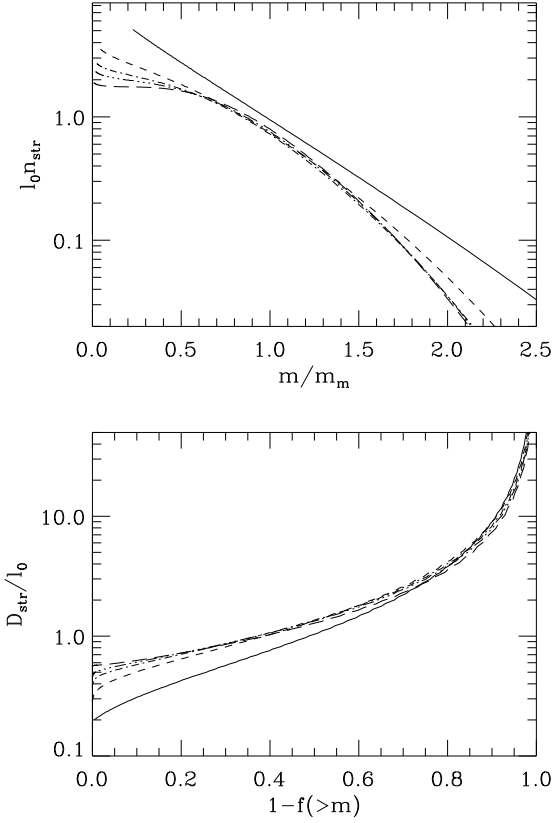}
\vspace{0.75cm}
\caption{Top panel: the mean comoving linear density of a pancake,
$l_0n_{str}(>m,\tau)$, vs. masses $m/m_m$ are plotted
for five moments of time: $\tau$ = 0.1 (solid line),
$\tau$ = 0.3 (dashed line), $\tau$ = 0.5 (dot-dashed line line),
$\tau$ = 0.7 (dot-dot-dot-dashed line), $\tau$ = 2 (long dashed
line). 
Bottom panel: the mean separations of pancakes, $D_{str}(>m,\tau)/l_0$,
vs. the fraction of compressed matter $f(>m)$, for
the same time moments. }
\end{figure}

\begin{figure}
\epsfxsize=7.5 cm
\epsfbox{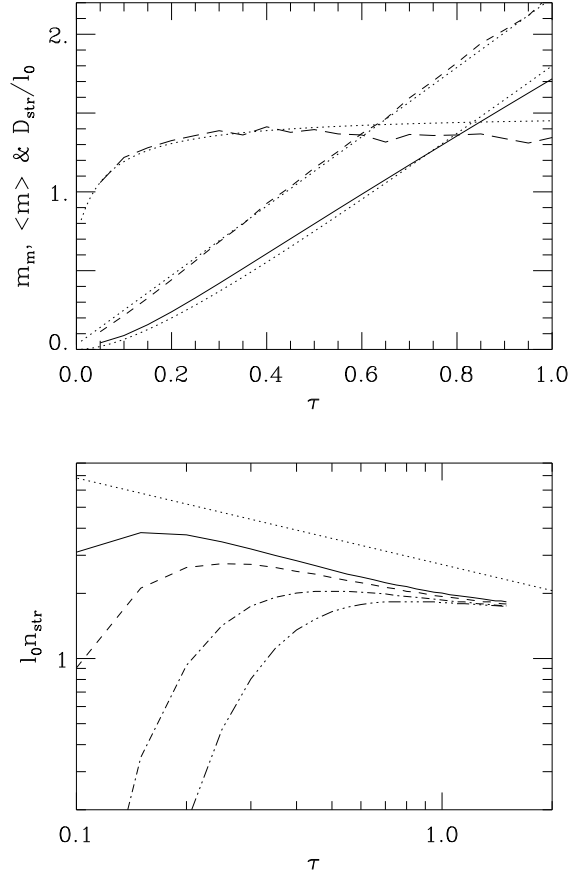}
\vspace{0.7cm}
\caption{Top panel: Time dependence of the mean mass, $<m>$,
(solid line), median mass, $m_m$, (dashed line) and the mean
separation $D_{sep}(m_m)/l_0$. The fits
(5.8) are plotted by dotted lines.  
Bottom panel: Evolution of the mean comoving linear number density
of structure elements, $n_{str}l_0$, vs. $\tau$ for four
threshold masses $m/l_0\geq$ 0.05 (solid line),
$m/l_0\geq$ 0.1 (dashed line), $m/l_0\geq$ 0.3 (dot-dashed line),
$m/l_0\geq$ 0.5 (dot-dot-dot-dashed line). A power law $\tau^{-0.4}$
is plotted by a dotted line. }
\end{figure}

Some of these functions are plotted in Figs. 1, 2 \& 3. Fig. 1
shows the PDF, $N_{str}(m,\tau)$, the mass function,
$mN_{str}(m,\tau)$, and the cumulative mass function, $f(>m)$,
for structure elements versus the $m/m_m$ for five
time moments. Here $m_m$ is the median mass of structure
elements defined by the condition $f(>m_m)=0.5$.
At small $\tau$  the distribution function is rapidly changing due
to the merging of smaller pancakes. Later
the evolution is slower and it is sustained by growth of median
mass due to the progressive matter concentration within richer
structure elements.

The same effect is clearly seen in Fig. 2 where the mean
linear number density of structure elements, $l_0n_{str}(>m,\tau)$,
is plotted for the same time moments. Rapid merging of low mass
pancakes and formation of more massive pancakes changes the
shape of this function at $m\ll m_m$ and $m\gg m_m$
while at $m\sim m_m$ evolution is very slow. The function
$D_{str}(f)$ plotted in Fig. 2 (bottom panel) is similar to that
found in simulations (DFGMM).

Time dependence of the mean, $<m>$, and median, $m_m$,
masses of structure elements is plotted in Fig. 3 (top panel)
together with fits
$$<m(\tau)>\approx 9\tau^2/(1+4\tau),\quad \tau\leq 1.5,
\eqno(5.8a)$$
$$m_m(\tau)\approx 0.03+2.2\tau, \quad \tau\leq 1.5.\eqno(5.8b)$$
Faster growth of $<m(\tau)>$ in comparison to (3.7) is
caused by merging of pancakes  as it is described by (5.5).
In Fig. 3 (top panel) the $D_{str}(m_m,\tau)$    
is plotted together with the fit
$$D_{str}(m_m,\tau)\approx 1.5\cdot l_0\sqrt{\tau\over \tau+
0.07}\eqno(5.8c)$$

Bottom panel in Fig. 3 shows evolution of the linear
number density of structure elements, $l_0n$, caused by their
exponentially fast formation at smaller $\tau$, and later, by
successive merging which is well fitted by the expression
$D_{str}(>q,\tau)\propto \tau^{-0.4}$ what is
slower then that found in DFGMM.

\subsection{Parameters of structure in observed catalogues}

In previous Sections the formation and evolution of structure
was described in a comoving space. But in observed catalogues the
redshift position of galaxies along the line of sight is used.
This difference distorts the parameters of observed structure
with respect to the theoretical expectations (see, e.g., Melott
et al. 1998; Hui, Kofman \& Shandarin 1999). To take into account 
this distortions the
theoretical relations given above need to be modified.

In observational catalogues the distance to a galaxy is defined 
by its observed velocity which can be found from (2.1) as 
$$v_i = {H(z)\over 1+z}[q_i-
\beta(z)B(z)S_i({\bf q})], \eqno(5.9)$$
$$\beta=1-{1+z\over B}{dB(z)\over d z},\quad 1\leq\beta\leq 2,$$
where $H(z)$ is the Hubble constant. Therefore, for observed
catalogues all relations obtained above will be valid after
replacement
$$\tau(z)\rightarrow \tau_v(z) = \tau(z)(\beta(z)\cos\varphi+
\sin\varphi),\eqno(5.10)$$
where $\varphi$ is a random angle between the line of sight
and the direction of pancake compression. It is evident that
$1\leq \tau_v/\tau\leq\sqrt{\beta^2+1}$, and for $\varphi=\pi/2
~~\tau_v=\tau$. This means that the observed structure parameters
will be randomly increased with respect to those found in the
comoving space.

The value of this distortion depends on the cosmological
parameters and redshift and is superposed with distortions
generated by the structure disruption, selection effects, and other
random factors. The well known example of such distortion is the
effect of so called ``finger of God''. The direct comparison of 
simulated
structure parameters in comoving and redshift spaces performed
in DMRT demonstrates a moderate dependence of the main structure
properties on these factors. But some properties of the structure
such as characteristics of galaxy distribution within walls are
different in comoving and redshift spaces. More
details can be found in Sec. 9 and DMRT.

\section{Formation of voids}

The same method can be used to solve the complementary
but more complicated problem of void formation. According to the
Zel'dovich theory the probability to find a 'void' with a size
$r>r_v$ is identical to the probability that the inequality
$$(1+z)r=q_1-q_2 - \tau\Delta S_{12}\geq (1+z)r_v\eqno(6.1)$$
holds under the conditions that pancakes have been  formed near
the boundary points $q_1 \& q_2$, and in the absence of any pancakes
between points $q_1 \& q_2$. This is equivalent to the probability
that the inequality
$$\Delta S_{12}\leq [q_{12}-(1+z)r_v]/\tau\eqno(6.2)$$
holds under the same conditions.

More accurately we can consider the {\it under dense regions}
bounded by pancakes with masses $m_1 \& m_2$ (or with masses
exceeding the threshold mass, $m_1, m_2\geq m_{thr}$) in the
absence of any pancakes with masses $m\geq m_{thr}$ between points
$q_1 \& q_2$. The methods discussed above and supplemented by
the void probability function technique could be used, in
principle, to obtain such probability but the derivation becomes
very cumbersome due to many additional conditions. Numerical methods
similar to that used by Sahni et al. (1994) could be more suitable
for such an investigation.

Let us remind however that the parameters of under dense regions are
closely linked to the spatial distribution of initial gravitational
potential what allows to specify conveniently  the large scale
perturbations. This link is clearly demonstrated by the adhesion approach
(see, e.g., Shandarin \& Zel'dovich 1989) and was discussed in details
in Paper I. It can also be described with the technique considered
above. Thus, for example, the large scale modulation of the pancake
distribution by the spatial distribution of this potential can be
described as a modulation of the effective 'time' moment $\tau_{eff}$.

To illustrate this statement we can consider the pancake formation
for a given potential difference, $\Delta \phi$, between two points
with comoving coordinates $x_1 ~\&~x_1+D_{cell}$. In this case the
distribution function of the pancake with a size $m\ll D_{cell}$
at the point with a comoving coordinate $x_1\leq y\leq x_1+D_{cell}$
is given by (3.6)~\&~(5.5) with
$${1\over\tau_{eff}(x)}\approx {1\over \tau}\left(1-\kappa_\phi
(D_{cell},y){\Delta\phi\over \sigma_\phi}\right),~ 0\leq y\leq
D_{cell},\eqno(6.3)$$
$$\sigma_\phi^2=2G_0(D_{cell}),\quad \kappa_\phi =
G_{12}(y) - G_{12}(D_{cell}-y).$$
Here $\sigma_\phi$ is the dispersion of gravitational potential
difference at the points with the separation $D_{cell}$. Functions
$G_0 \& G_{12}$ are introduced in Appendix I.

As is seen from (6.3) the efficiency of  pancake formation
depends directly on the coordinate $y$ and the  potential
difference, $\Delta\phi(D_{cell})$. It is amplified in regions with
$\kappa_\phi\Delta\phi\leq$0, and depressed in regions with
$\kappa_\phi\Delta\phi\geq$0.

The same large scale modulation modifies the Euler position of
pancake and instead of Eq. (3.16) we obtain after averaging
of $\Delta \phi(m)/m$ (under the condition $m\ll D_{cell}$)
$$<(1+z)r_{pan}> = y +\tau G_r(y,D_{cell})
{\Delta\phi\over \sigma_\phi},\eqno(6.4)$$
$$G_r(y,D)=yG_{12}(y)+ (D-y)G_{12}(D-y),$$
and for the mean Eulerian linear number density of pancakes
$$n_e(y) = \left({\partial r_{pan}\over
\partial y}\right)^{-1} = {1+z\over 1+\tau
G_n(y,D_{cell}){\Delta\phi\over \sigma_\phi}},\eqno(6.5)$$
$$G_n(y,D)=G_{12}(y)-G_{12}(D-y)+y^2 G_{23}(y)-(D-y)^2G_{23}(D-y).$$

These expressions describe the matter infall into wells of initial
gravitational potential and clearly demonstrate the close connection
of under dense regions with the spatial distribution of initial
gravitational potential. More detailed discussion of this effect
can be found in Paper I.

\section{Large scale bias and characteristics of biased cold matter
distribution}

\subsection{Large scale bias}

In observations we have to deal with the galaxy distribution
which is, probably, biased relative to the spatial distribution
of dark matter in the considered range of scales. The qualitative
physical ground of such large scale bias is transparent and was
discussed already ten years ago by Dekel \& Silk (1986), and Dekel
\& Rees (1987). It can be caused by the joint action of large scale
perturbations and reheating of the baryonic component of matter.
The interaction of large and small scale perturbations shows
up as an excess of low mass pancakes and
acceleration of their growth within richer wall-like elements
and is seen in simulations as a strong concentration of high
density peaks within filaments and richer walls.
It results in an excess of compressed cold gas  inside
"proto" walls before reheating. But after  reheating further
compression and cooling of gas are inhibited due to the growth of
entropy of uncompressed gas. This excess of {\it cold, low entropy
baryonic matter}, reached at the reheating redshift, $z=z_h$,
can be  considered as the excess of galaxies within
richer walls that is the large scale bias in spatial galaxy
distribution.

In Paper I the modulation of the rate of pancake formation by the
large scale perturbations of initial gravitational potential
was discussed. Here we consider the direct interaction of
earlier and later pancakes, that allows us to obtain more
reliable estimates of the large scale bias. With the technique
developed above we can consider only the direct interaction of
two population of pancakes, namely, those created
at $z=z_h$, and at $z=0$, and we have to neglect the mutual
influence of intermediate population of pancakes accumulated by
the walls. Such an influence can in principle be considered but
the description becomes very cumbersome. So, our estimate presented
below is actually the lower limit of the possible large scale bias.
Perhaps, this problem can be studied in simulations
which take into account all important factors together (see, e.g.,
Sahni et al. 1994; Cole et al. 1998).

For the warm dark matter model discussed in Paper I the mass of
DM particles restricts the minimal size of created
pancakes to the correlation scale of initial density perturbations,
$r_c$. For the CDM power spectrum considered here there are no
natural limits and, in principle, pancakes with arbitrarily low
mass can form. Formally, it is due to the divergence of higher
moments of power spectrum and, because of this, the typical scale
$r_c$ is zero. In this case, however, the typical scale $q_0$
introduced by (3.4) plays similar role and discriminates the
process of earlier low mass pancake formation, described through
the deformation tensor, and later formation of structure
elements discussed above. The typical mass of such low mass pancakes
given by the expression (3.4) exceeds the
estimates of minimal baryonic mass formed before the reheating,
$M_{min}\sim 10^6M_\odot$ (Tegmark et al. 1997), and the value $q_0$
can be accepted as a natural limit of pancake size. Further
evolutionary history of low mass pancakes is uncertain, and this
problem should be considered separately.

Thus, for the CDM power spectrum we assume that:
\begin{enumerate}
\item All baryons accumulated by structure elements with a size
$m\geq q_0$, at the  redshift $z = z_h, ~\tau=\tau_h$,
are incorporated into observed 'galaxies'.
\item The formation of such clouds is interrupted at the
redshift $z=z_h,~\tau=\tau_h$ by the instantaneous reheating of the
uncompressed gas.
\end{enumerate}
Both assumptions are very restrictive but they can be
used for rough estimates of the efficiency of this mechanism of
generation of the large scale bias.
The moment of reheating is, probably, restricted to $7\leq z_h\leq$
15. The lower limit is imposed by the observational constrains. The
upper limit is imposed by the very fast growth of the inverse Compton
cooling for larger redshifts.

This means that for the models with  $\Omega_m~\sim$ 0.3 -- 1 the
reheating probably occurred at $\tau_h = \tau(z_h)\approx$
(0.15 -- 0.3)$\tau_0$.

\subsection{Statistical characteristics of biased cold matter
distribution}

The biasing factor on the LSS and SLSS scales can be estimated in
the same manner as the probability of merging (5.2).
Now we are interested in the mass of cold matter collapsed at
$\tau=\tau_h$ which was accumulated at the moment $\tau$ by a
pancake with a size $D\gg m_{cld}$. The approximate PDF for this
mass can be written as
$$N_{cld}(m_{cld},\tau_h)\approx
2 N_{c}\left(m_{cld},\tau_h;D,\tau,{m_{cld}+D\over 2}\right)$$
$$\approx N_{cr}(m_{cld},\tau_1),\eqno(7.1)$$
$${1\over \tau_1} = {1\over \tau_h} - {\kappa_1(D)\over \tau},\quad
\kappa_1(D)\approx {1\over (2+D)(1+D)}.$$
These relations show that, as it was discussed in Paper I, the
formation of earlier pancakes within larger structure elements
is accelerated, what is seen in (7.1) as a shift of the
effective evolutionary 'time', $\tau_1$, relatively to
the real reheating 'time' moment, $\tau_h$. The corresponding
mean 'mass' of such pancakes can be found as
$$<m_{cld}(\tau_h)>\approx 4\tau_1^2\approx
4\tau_h^2(1+1.5\tau_h/\tau),\quad D< 1.\eqno(7.2)$$
The dimensionless biasing factor
$$b_{bias} = {<m_{cold}>\over <m>}-1\approx 1.5\tau_h/\tau,
\eqno(7.3)$$
is about of 1/5 -- 2/5 and weakly depends on the size $D$ of
the considered structure element.
The relation (7.1) describes
also the excess of larger pancakes induced by the
influence of larger structure elements
$${N_{cld}(m_{cld},\tau_h)\over N_{cr}(\mu,\tau_h)}\approx
{\tau_h\over \tau_1}
\exp\left({m_{cld}\over 2\tau_h\tau}\right).\eqno(7.4)$$
As is seen from (7.1) the bias decreases, for $\tau\gg \tau_h$.
This means that the indirect multi step interaction becomes
specially important and the intermediate population of
pancakes formed at $\tau_h\leq \tau^*\leq \tau$ with a size
$x^*\leq D$ and accumulated by a structure element with
a size $D$, at the moment $\tau$, will enhance the bias and
increase the concentration of cold matter within larger
pancakes.

These results are consistent with the conclusions of Paper I, Cole
et al. (1998) and Sahni et al. (1994) and demonstrate the sensitivity
of the considered bias to the detailed characteristics of the process
of galaxy formation. They show that the over and under dense regions
found now in the observed galaxy distribution can, probably, be
related even at high redshifts to a spatial modulation of sizes of
the earlier pancakes. The larger pancakes are more closely associated
with over dense regions whereas the under dense regions are occupied
preferentially by low mass pancakes.

Summarizing, we can infer that the influence of large scale bias
generated by a combined action of reheating and large scale
perturbations can be important and can distort characteristics of
the observed galaxy distribution relative to the same
characteristics of DM distribution. Under suitable conditions the
fraction of baryons accumulated by galaxies  can be suppressed
in low massive pancakes and under dense regions and increased in
richer structure elements.

\section{Dynamical characteristics of the pancakes}

In this section we will find two dynamical characteristics of
pancakes, namely,  the velocity of pancakes as a whole and the
dispersion of matter velocity within pancakes caused by
compression of  pancakes.

The velocity of an infalling particle, $v_i$, can be found from (2.1) as
$$v_i({\bf q},z) =dr_i/dt = {H(z)\over 1+z}[q_i-
\beta(z)B(z)S_i({\bf q})], \eqno(8.1)$$
$$\beta=1-{1+z\over B}{dB(z)\over d z},$$
where $H(z)$ is the Hubble constant.

The functions $\beta(z)$ can be found from (2.2) and (2.3) as
$$\beta\approx 2-B^3(z){1-\Omega_m\over
1+1.2\Omega_m},\quad \Omega_m+\Omega_\Lambda=1, \eqno(8.2)$$
$$\beta\approx 2-B(z){1-\Omega_m\over1+1.5\Omega_m},
\quad \Omega_m\leq 1,~~~ \Omega_\Lambda=0.\eqno(8.3)$$
For $\Omega_m=1$ both expressions give
$\beta = 2$.

Using the general relation for the velocity of a particle
(8.1) we can find the velocity of a pancake as a whole through
the momentum conservation law, we have
$$v_{pan} = {1\over q_{12}}\int_{q_2}^{q_1}v(q)dq =
{H(z)\over 1+z}\left[q_{cent}-\beta B(z){\Delta\phi_{12}
\over q_{12}}\right],$$
$$\Delta\phi_{12}=\phi(q_1)-\phi(q_2),~~
q_{cent}=(q_1+q_2)/2.\eqno(8.4)$$
The pancake position, $r_{pan}$, was given by (3.16).
The peculiar velocity of pancake is
$$u=v_{pan}-H(z)r_{pan} = {H(z)\over 1+z}B(\beta-1)
{\Delta\phi_{12}\over q_{12}}.\eqno(8.5)$$
The mean peculiar velocity, $<u> = 0$, and its dispersion,
$\sigma_{u}$, is
$$\sigma_{u}^2(z) = {H^2(z)\over (1+z)^2}[\beta(z)-1]^2
B^2(z){2G_0(q_{12})\over q_{12}^2}.\eqno(8.6)$$
The distribution function of peculiar velocity is Gaussian as
before. In dimensional variables, at $z=0$, and for $q_{12}\ll 1$,
we have:
$$\sigma_{u}(0) = H_0[\beta(0)-1]\sigma_s/\sqrt{3}=
H_0[\beta(0)-1]l_0\tau_0.\eqno(8.7)$$
These relations show that for smaller $q_{12}$ the pancakes
velocity is close to the mean velocity dispersion and decrease
$\propto q_{12}^{-1/2}$, for $q_{12}> 1$. The vortexless character
of the initial velocity perturbations is clearly seen in eq. (8.4),
(8.5).

Variations of the pancake velocity with the transversal coordinates
can distort the shape of pancake and, in principle, can destroy the
pancake. The impact of this factor can be characterized through the
expression (8.4) as a correlation of pancake velocity at different
transversal coordinates. This correlation is described  by the
normalized function $r_{v}$ (see Appendix IV), and for
$q_{12}, q_{34}\ll$ 1, we have
$$r_v = {\bf q}_{12}\cdot{\bf q}_{34}/q_{12}q_{34},\eqno(8.8)$$
what demonstrates the regular character of velocity variations. For
larger $q_{12}$  and $q_{34}$ the velocity dispersion (8.6) decreases
and therefore this factor becomes less important.

The velocity dispersion within a pancake with a given mass
$q=q_{1}-q_{2}$, caused by the matter infall into pancake, is defined
by the energy conservation law. The velocity of a particle with the
coordinate $q_3$ relative to the pancake as a whole is
$$v_{inn}(q_3,z) = v(q_3,z)-u(z)=\eqno(8.9)$$
$$ {H(z)\over 1+z}\left[q_3-q_{cent}-
\beta(z)B(z)\left(S(q_3)-{\Delta\phi_{12}\over
q_{12}}\right)\right].$$
Using the conditional mean values and dispersions of random
functions $S(q_3)$, and $\phi(q_1)-\phi(q_2)$, listed in Appendix I,
we obtain the required velocity dispersion
$${(1+z)^2\over H^2(z)}\sigma_{v}^2 =
{q^2\over 12}[(\beta(z)-1)^2+2\beta(z)f_1(q)- $$
$$\beta^2(z)f_2(q)+\tau^2\beta^2(z)f_3(q)].\eqno(8.10)$$
The functions $f_1,~~f_2,~\&~ f_3$ are given in Appendix IV.
The velocity dispersion $\sigma_v(q)$ is plotted in Fig. 4 for a
few cosmological models, for $z=0$, $h=0.7$.

\begin{figure}
\centering
\epsfxsize=8 cm
\epsfbox{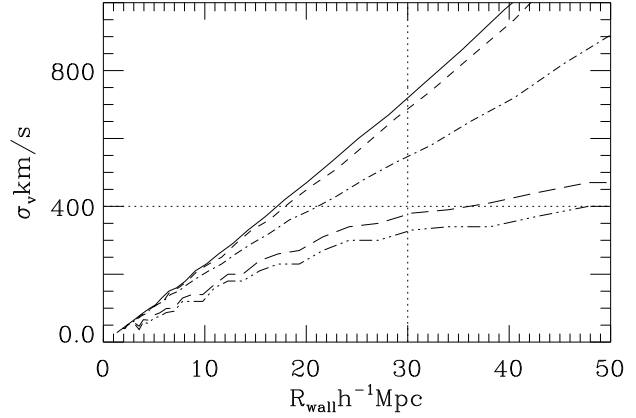}
\vspace{0.5cm}
\caption{The velocity dispersion, $\sigma_v$, vs. the size of
"proto walls", $q=R_{wall}$, for the redshift $z=0$, for five
cosmological models with $h=0.5$, $\Omega_m=1, \Omega_\Lambda=0$
(solid line),  and $h=0.7$,
$\Omega_m=0.5, ~\Omega_\Lambda=0.5$ (dashed line),
$\Omega_m=0.5, ~\Omega_\Lambda=0$ (dot-dashed line),
$\Omega_m=0.35, ~\Omega_\Lambda=0.65$ (dot-dot-dot-dashed line),
$\Omega_m=0.3, ~\Omega_\Lambda=0$ (long-dashed line).
Dotted lines show  observational estimates of the size of "proto
walls" $R_{wall}\sim 30h^{-1}$Mpc, and the velocity dispersion
$\sigma_v\sim$ 400km/s.}
\end{figure}

For $q_{12}\leq 1$ the functions
$f_1, f_2 \& f_3$ are small, $f_1\leq 0.06,~~ f_2\leq 0.13$, and
$f_3\leq 0.2$, and the contribution of the three last terms in
(8.9) becomes important only when $\beta\sim 1$ that is for
cosmological models with a lower matter density at redshift
$z\leq 1$. For poorer pancakes when the first term dominates,
the simple approximate relation
$$\sigma_{v}\approx H(z)(1+z)^{-1}(\beta-1){q_{12}\over
2\sqrt{3}}\eqno(8.11)$$
can be used for estimates of the expected velocity dispersion.
 For such pancakes the distribution function of $\sigma_{v}$ is
similar to (4.5).

\section{Comparison with simulated and observed structure parameters}

Some characteristics of large scale galaxy distribution can be
extracted from available catalogues of galaxies and clusters
of galaxies and catalogues of absorbers in spectra of
quasars. Some characteristics of large scale DM distribution can be
derived from available simulations. All these characteristics
are distorted by selection effects, bias between spatial distribution 
of DM and galaxies, and other factors what complicates the direct
comparison of characteristics obtained with different methods and 
data bases.

Nonetheless, the comparison of even so distorted characteristics
with the expected characteristics discussed above can be interesting
as it allows one to compare various estimates and illustrates their
sensitivity to basic cosmological parameters and other factors. More
detailed discussion can be found in DMRT ~\&~LCRS2.

\subsection{Observed and simulated parameters of the structure}

Analysis of galaxy distribution in the Las Campanas Redshift Survey
(Shectman et al. 1996) shows that the richer and poorer structure
elements can be assigned to wall-like and filamentary populations,
but the accurate demarcation of these subpopulations is problematic.
Both subpopulations accumulate $\approx$ 40 -- 60\%
of galaxies. Filaments fill the gaps between the walls and form a
random network with the mean cell size $D_f^{obs}\sim$~10 -- 12$h^{-1}$
Mpc at $z=0$. The richer walls can be formed due to an
anisotropic compression of matter within slices with
a typical size of $R_{wall}^{obs}\approx$ 20 -- 25$h^{-1}$Mpc and
with a typical separation of $D_{wall}^{obs}\sim$ 50 -- 60$h^{-1}$Mpc
(LCRS1 \& LCRS2). The walls are significantly disrupted by the small
scale clustering of galaxies (see, e.g., Fig. 5 in Ramella et al. 1992).
The velocity dispersion within such wall-like elements was estimated
as $\sigma_v^{(obs)}\sim~$ 350 -- 400 km/s (Oort 1983a). The bulk
velocities of galaxies are now estimated as $\sigma_u^{obs}\sim$ 
400km/s (see, e.g., Dekel 1997). The observed surface density of 
structure elements is heavily distorted by the selection effects and 
small scale clustering and is equally well fitted to a power law, 
exponential distribution, and functions discussed in Secs. 4~\&~5.

The structure in the DM distribution was simulated and analyzed by DMRT
for the SCDM model with $\Omega_m$=1, $h$= 0.5, $\Lambda$CDM model with
$\Omega_m$=0.35, $\Omega_{\Lambda}$=0.65, $h$ = 0.7 and OCDM model with
$\Omega_m$=0.5, $h$ = 0.6. For all these models the normalization to the
two year COBE data were used. These simulations reproduce the structure
with the main parameters similar to those
found in observations. The analysis shows that walls accumulates
$\sim$ 40\% of DM with a mean size of "proto walls"
$R_{wall}^{sim}\sim$ 15 -- 20$h^{-1}$Mpc and with a mean separation
of walls and filaments $D_{wall}^{sim}\sim$ 50 -- 70$h^{-1}$Mpc,
$D_{f}^{sim}\sim$9 -- 14$h^{-1}$Mpc, respectively.

In all the models the simulated velocity dispersion within walls,
$\sigma_v^{sim}$, is the same along all three principal directions
of walls and it is generated by the wall curvature and disruption into
a system of high density clumps. The walls are seen usually as ordered
sets of irregular high density clumps connected by lower density bridges.
The degree of wall disruption depends on the code used and reached
resolution, and it is more conspicuous in the models with larger 
$\Omega_m$. In the redshift space this inner structure is partly 
eroded and characteristics of DM walls are similar to that found for 
the observed galaxy distribution.

The isotropy of simulated velocity dispersion $\sigma_{v}$ confirms
the essential influence of small scale clustering on the properties
of both observed and simulated structure elements. The instability of 
a thin compressed layer was analyzed in the linear approximation by
Doroshkevich (1980), and Vishniac (1983), and was recently
simulated by Valinia et al. (1997). The clustering rate and
parameters of formed clusters depend on the density of compressed
matter and properties of transversal velocity field.

\subsection{Comparison of simulated and expected structure properties}

\begin{table*}
\begin{minipage}{180mm}
\caption{Main expected parameters of the SLSS.}
\label{tbl2}
\begin{tabular}{cccccccccccc} 
$\Omega_m$&$\Omega_\Lambda$&h&$l_0$&$\tau_0$
&$R_{wall}$&$D_{str}$&$\tau_u$&$m_v/l_0$&$\tau_v$&$m_p/l_0$&$\tau_p$\cr
$$&$$& &$h^{-1}$Mpc& &$h^{-1}$Mpc&$h^{-1}$Mpc& & &\cr
1~~~~&0~~~~ &0.5~~& 13.2&0.7&$\sim 20.7$&$\sim$19.8&0.52&1.2&0.42&1.0&0.41\cr
0.5~~&0.5~~ &0.7~~& 18.9&0.6&$\sim 25.5$&$\sim$26.8&    &   &    &   &    \cr
0.5~~&0~~~~ &0.6~~& 22.0&0.3&$\sim 14.5$&$\sim$26.9&0.28&0.5&0.27&0.8&0.35\cr
0.35 &0.65  &0.7~~& 26.9&0.4&$\sim 23.5$&$\sim$37.1&0.40&0.9&0.36&1.3&0.44\cr
0.35 &0~~~~ &0.7~~& 26.9&0.2&$\sim 12.0$&$\sim$34.5&    &   &    &   &    \cr
\end{tabular}
\end{minipage}
\end{table*}

Theoretical parameters of DM wall-like structure elements are listed
in Table I for the median mass $m=m_m$ and for $l_c/l_0=$0.056.
The variations of $R_{wall}$ listed in Table 1 are moderate and,
in the range of precision of our approximate consideration,
these values are similar to those observed and simulated. The
differences become more significant only for the OCDM models.
The value of $D_{str}$ is sensitive to the assumed mass threshold
(see Figs. 2 \& 6) and can be adjusted.

The simulated structure elements are also distorted by the small
scale clustering but owing to the rich statistics of such elements
a more detailed quantitative comparison of simulated and expected
properties of the structure can be performed. To do this we will
consider the PDFs (4.4) and (4.5) for the surface density of filaments
and pancakes, the PDFs for velocity $u$ and velocity dispersion
$\sigma_v$ and direct estimates of important parameters
$$\tau_u = {\sqrt{<\sigma_u^2>}\over H_0l_0(\beta-1)},\quad
m_v/l_0 = {2\sqrt{3<\sigma_v^2>}\over H_0l_0(\beta-1)},\eqno(9.1)$$
following from relations (8.7) and (8.11).

\begin{figure}
\centering
\epsfxsize=8 cm
\epsfbox{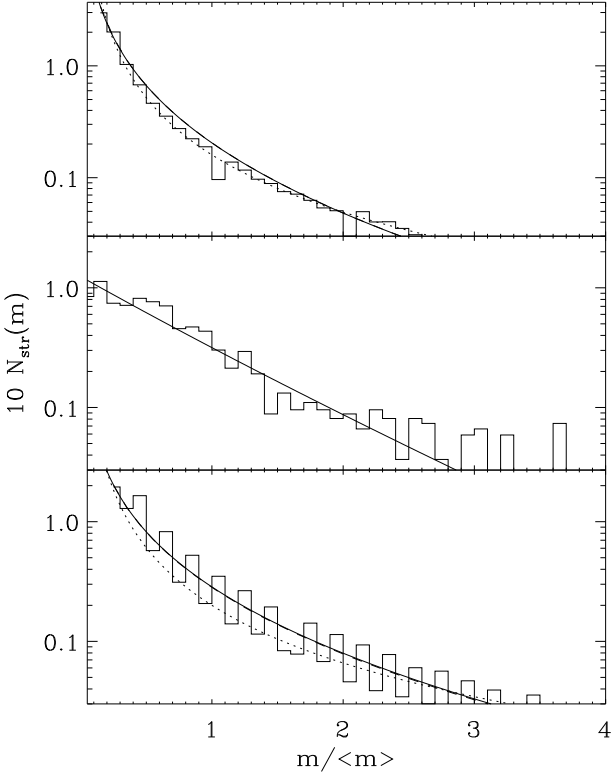}
\vspace{0.8cm}
\caption{The simulated PDFs of surface density of structure elements,
$N_{str}(m)$, vs. $m/<m>$ for the full sample (top panel) and subsamples
of richer (middle panel) and poorer (bottom panel) structure elements
for the $\Lambda$CDM model. The fits (9.3) and (9.4) are plotted by
solid lines, the power fits are plotted by dashed lines.}
\end{figure}

\begin{figure}
\centering
\epsfxsize=8 cm
\epsfbox{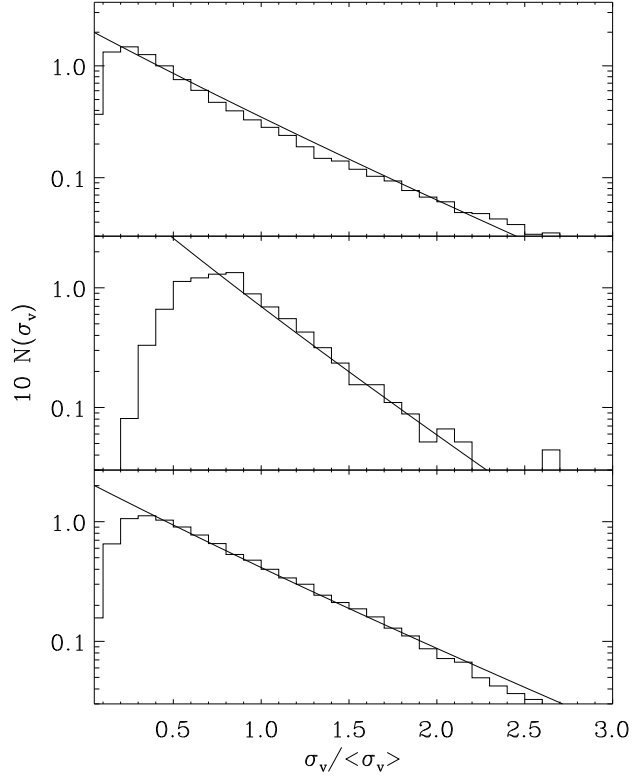}
\vspace{0.8cm}
\caption{The simulated PDFs of velocity dispersion of structure elements,
$N_{str}(\sigma_v)$, vs. $\sigma_v/<\sigma_v>$ for the full sample
(top panel) and subsamples of richer (middle panel) and poorer
(bottom panel) structure elements for the $\Lambda$CDM model. The fits
(9.3) are plotted by solid lines.}
\end{figure}

To characterize the filamentary and wall-like subpopulations of
structure separately the analysis was performed  both in the comoving 
and redshift spaces for three subsamples of structure elements.
The first contained all particles, the second and third incorporated
the richer and poorer structure elements with $N_{mem}>$ 200 and
$N_{mem}<$ 200, respectively. Here $N_{mem}$ is the number of DM
particles within a structure element bounded by the threshold density
$n_{thr}/<n>$=1.7, 1.1~\&~1.1 for the SCDM, OCDM and $\Lambda$CDM models.
In all the cases the second subsamples accumulates $\sim$40 -- 45\% of
particles.

To obtain the required PDF a set of rectangular sampling cores was prepared
and the number of DM particles in the intersection of separate structure
elements with these cores
was taken as a characteristic of surface density of structure elements.
For each of the cluster its velocity $u$ and velocity dispersion
$\sigma_v$ along the core were also
found. To depress the impact of small scale clustering the analysis was
performed for the core size $L_{core}=10h^{-1}$Mpc, for the second
subsample, and for $L_{core}=4h^{-1}$Mpc, for the first and third 
subsamples when the core size is restricted by the separation of 
filaments. The random intersection of cores and structure elements 
generates significant excess of poorer clusters. To depress this 
effect poorer clusters were rejected and the truncated PDFs were 
considered. Even so, the number of clusters used was $\sim$ 3 500, 
for the second, and $\sim$ 10 000, for the first and third subsamples. 
The random orientation of structure elements and cores can increase the
measured surface density (up to two times for homogeneous matter
distribution, Kendall~\&~Moran 1963), and 
reduces the measured velocity of structure elements along the core, 
$u_c$, by a factor of $\sqrt{3}$ and we have 
$$u_c = u\cos\varphi,\quad \tau_u = \sqrt{3}\tau_{uc}.\eqno(9.2)$$

To compare the simulated and theoretical PDFs the two parameters fits
were used. For the second subsample the distribution of surface density
was fitted to the expression
$$N_w = {a_w\over \sqrt{x_w}}e^{-x_w}erf(\sqrt{x_w}),
\quad x_w={b_w m_p\over <m_p>},\eqno(9.3)$$
which reproduces approximately the theoretical relations
(4.5) with the correction to the pancake merging discussed in Sec. 5.
For the first and third subsamples the expression
$$N_f = {a_f\over x_f^{1.7}}e^{-x_f}[1+erf(\sqrt{x_f})],\quad
x_f=\sqrt{b_f m_f\over <m_f>},\eqno(9.4)$$
reproduces approximately the theoretical relations (4.4) and gives 
better fits. Here parameters $b_w~\&~b_f$ describe the truncated 
character of fitted PDF whereas parameters $a_w~\&~a_f$ provide its 
normalization. For velocity of clusters, $u$, the PDF is found to 
be well fitted to the expected Gaussian distribution, the expression 
(9.3) with parameters $a_v~\&~b_v$ fits the velocity dispersion 
$\sigma_v$.

For $\Lambda$CDM model these PDFs are plotted in Figs. 5 and 6 together
with the best fits. It turns out that different PDFs verify the accepted
discrimination of two populations of structure elements. For the second
subsample the theoretical estimate of mean wall size
$$<m_p>/l_0\approx 8(0.5+1/\pi)\tau_0^2\approx 6.55\tau^2_0,\eqno(9.5)$$
links the parameters $\tau_0$ and $<m_p>$. For the $\Lambda$CDM and
OCDM models the parameters $\tau_u$ (for all samples) and $<m_v/l_0>,
~\tau_v=\sqrt{<m_v/l_0>/6.55}, 
~<m_p/l_0>~\&~\tau_p=\sqrt{<m_p/l_0>/6.55}$
(for second subsample) listed in Table I are consistent with the 
simulated amplitude. For the first and third subsamples the 
distributions plotted in Fig. 6 and corresponding parameter 
$<m_v/l_0>$ characterize the typical mass of dominant numerous poorer 
elements and it is  about half of that for the 
second subsample. 

The formal precision of these estimates can be 
taken as $\sim$ 7 -- 10\% (precision of fits) but a real precision 
can be estimated by the comparison with the value of the input parameter
$\tau_0$ also listed in Table I. 
Differences between results obtained in comoving and redshift
spaces do not exceed the reached precision. 
For the SCDM model the expected and reconstructed parameters 
differ by a factor of $\sim$ 1.5 what can
be partly caused by the strong wall disruption. Moreover, this 
simple description is correct for $m/l_0\leq$1 whereas more cumbersome
general relations describe the properties of richer walls which dominate 
the SCDM model.

In all these cases the simulated mass distribution can be equally well
fitted to a power law with the exponent $\kappa\sim$ 1.5 -- 2. Such
distribution is similar to that described analytically for the matter
concentration within a set of clusters with a surface density
$\sigma_{cls}\propto r^{-\gamma}$ (see also discussion in Miralda-Escude
et al. 1996). In this case the measured PDF and mass function are also
expressed by the power law
$$dW_{cls}\propto \sigma_{cls}rdr\propto 
\sigma_{cls}^{-2/\gamma}d\sigma_{cls}.\eqno(9.6)$$
Of course, profiles of separate irregular clusters can vary over a wide
range and such description and interpretation illustrate only the
important role of small scale clustering.

\subsection{Evolution of DM structure }

Results obtained in Secs. 3, 4 \& 5 show that during the most
interesting period $\tau\gg l_c/l_0$ evolution of DM structure shows
features of self-similarity and important characteristics of
structure can be expressed as functions of the parameter $\mu(q)/\tau$.
This approximate self-similarity is caused by the Zel'dovich
approximation and occurs for any distribution function
of initial perturbations and any power spectrum. It becomes more 
transparent for simple  structure functions, at 
$q_0\ll q\leq$1 (Appendix I), typical for the CDM like initial power 
spectrum, but becomes more cumbersome for $\tau\sim
l_c/l_0\ll$ 1, and $\tau\geq$ 1, when the influence of the scales
$q_0 \& L_0$, introduced by (I.6), (I.7) becomes important. This
self-similarity allows us to characterize the structure evolution 
by the time dependence of several typical parameters such as the 
mean and median masses, $<m>~\&~m_m$.

This approximate self-similarity is partly violated due to the
evolution of pancakes, discussed in Sec. 4, and small scale pancake
disruption as they are not described by the Zel'dovich theory. 
Nonetheless, the observations and simulations show that in a universe 
dominated by cold DM particles general properties of matter
distribution are quite similar at high and small redshifts (see more
detailed discussion in Governato et al. 1998; Jenkins et al. 1998).
Self-similarity of structure evolution was previously discussed for 
the scale-free power spectra (see, e.g., Efstathiou et al. 1988).

The evolution of observed structure formed by galaxies can however
be far from the self-similar evolution of DM structure due to the
small and large scale bias. Nonetheless, as it was discussed above,
the main parameters of DM structure can be compared with observed
galaxy distribution. Thus, in particular, the observed separation 
of filaments, at $z=0$, is  about 4 -- 6 times smaller  then the 
typical separation of walls (see, e.g., Efstathiou et al. 1988).
Such result can be
attributed to pancakes formed by compression of a slice with
thickness $m_f\sim$ (0.1 -- 0.05)$m_m\approx$ 1 -- 2$h^{-1}$Mpc at
redshift $z\sim$ 3 -- 5. In smaller pancakes the formation of galaxies
can be suppressed and they can be associated with weak Ly-$\alpha$
absorbers observed far from galaxies (Morris et al. 1993; Stocke et
al. 1995; Shull et al. 1996).

\section{Summary and Discussion}

In this paper we continue investigations and statistical description
of the process of structure formation and evolution initiated
 in our previous papers
(Buryak et al. 1992; Demia\'nski \& Doroshkevich 1992, 1997;
DFGMM; Paper I) and based on the nonlinear theory of gravitational
instability (Zel'dovich 1970). The new elements discussed above are
the approximate statistical description of the nonlinear structure
evolution manifesting itself as a successive
matter concentration into more and more massive structure elements
and the estimates of the large scale modulation of the spatial
distribution of luminous matter relative to DM and baryonic
components. We show that, as has been discussed in Sec. 9.4,
the evolution of DM structure demonstrates some features of
self-similarity and the main characteristics of the DM structure
can be expressed through the structure functions of
initial perturbations and through parameters of the initial
power spectrum and cosmological models. It is shown -- in accordance
with simulations -- that in a low density models the nonlinear evolution
occurs on the scale $\sim$ 20 -- 25$h^{-1}$Mpc and results in
formation of wall-like component of the structure of the universe.

Results discussed in Section 3 show that, for the CDM-like power
spectra, the high nonlinear matter compression into low mass pancakes
and the percolation take place already at $\tau\approx l_c/l_0\ll$1.
Later evolution leads to a rapid growth of typical mass of DM
pancakes. The theoretical description
discussed above can be directly applied to the interpretation of deep
pencil beam galactic surveys and absorption spectra of quasars, what
allows us to consider together rich observational data accumulated
by these methods for both small and large redshifts. For larger 3D
catalogs and simulations this description can be applied to
results  obtained with the core-sampling analysis as it was 
demonstrated in Sec. 9.2 . 

Comparison with simulations shows that the velocity dispersion
of matter compressed within the wall-like elements is more sensitive
to the cosmological parameters, large scale modulation of spatial
galaxy distribution and properties of DM component. The small scale
clustering generates the essential difference between the expected
and simulated velocity dispersion and allows us to discriminate the
cosmological models with respect to the matter density, $\Omega_m$,
and the composition of DM. This clustering is smaller for the models
with $\Omega_m\leq 0.5$, but for models with $\Omega_m\leq 0.3$ it
is more difficult to reproduce the sizes, separations of observed
walls, and the high matter concentration within the SLSS. The 
observed structure parameters are best reproduced for 
the models with $\Omega_m h\approx$ 0.2 -- 0.3. These values are  
consistent with estimates obtained both from the simulations of 
cluster of galaxies (Bahcall \& Fan 1998; Cole et al. 1997, 1998) 
and from the observations of high redshift supernovae (Perlemutter 
et al. 1998). 

Two factors can extend the set of acceptable cosmological parameters.
The first is a more complicated composition of DM component, what
means that an essential fraction of DM  can be associated with
relatively hot and/or low mass particles (see, e.g., Colombi et al.
1996; Brustein \& Hadad 1998). The simple estimates based on the 
Trimain - Gunn relation
(Trimain \& Gunn 1979) show that if the recently formed wall-like
SLSS is sensitive to the influence of low mass relic particles, with
$M_p\leq$ 3 -- 5 eV, then the structure properties at higher redshifts
are sensitive to more massive particles as well.  This means that for
the MDM models with more complicated DM composition the agreement
between observed and simulated properties of the LSS and SLSS can
be achieved in a broader range of $\Omega_m$ and $h$ and, moreover,
the analysis of observed structure evolution can specify the DM
composition. The models including some fraction of unstable
dark matter particles (Turner et al. 1984; Doroshkevich et al. 1989)
seem also to be perspective. Unstable particles can produce reheating 
by photo decay whereas the hot products of decay will delay both
formation and disruption of the walls.

The second factor is the large scale modulation of spatial galaxy
distribution. The formation of LSS and SLSS is always accompanied by
bias on the same scales, because both processes are caused by the
same large scale perturbations and, so, are strongly correlated in
space. The efficiency of such bias depends on many factors, such as
the redshift of reheating period, and the formation and evolution of
cold clouds. The rough estimates of the bias presented in Sec. 7
show that the bias factor could be quite high (see also Demia\'nski
and Doroshkevich 1997; Paper I) and together with other factors
it can essentially distort the spatial distribution of galaxies with
respect to the DM distribution. These estimates could
be enhanced by taking into account the mutual interaction of pancakes
with various sizes and moments of formation. More representative
estimates of the bias can, probably, be obtained from simulations
such as, for example, Sahni et al. (1994) and Cole et al. (1998).

This mechanism of bias generation implies early, for $z\geq 5$,
formation of the main fraction of low entropy gaseous clouds, that
can be identified with 'proto galaxies'.
The reheating does not prevent further formation of DM pancakes, which
can be identified with a population of gas clouds responsible for
weaker absorption lines
observed at high redshifts (Miralda-Escude et al. 1996; Hernquist et
al. 1996) but it inhibits the formation of high density, low entropy
gaseous clouds and delays the formation of galaxies in extended under
dense regions. Further transformation of  formed earlier
'proto galaxies' into observed galaxies is a slow and complicated
process continuing up to now and earlier formation of such clouds
is not in contradiction with the observed peak of galaxy formation
at redshifts $z\approx$ 2 -- 3 (Steidel et al. 1996).

The observed strong variations of galaxy distribution
with respect to the DM and intergalactic gas distribution,
$\rho_{gal}/\rho_{gas}$, can be, partly, associated with this
bias. Indeed, if in clusters of galaxies this ratio is found to be
$\rho_{gal}/\rho_{gas}\approx 0.2$ (see, e.g., White et al. 1993)
then, for example, within Bo\"ots Void
$\rho_{gal}/\rho_{gas}\rightarrow 0$ (Weistrop et al. 1992). The
existence of 'invisible' structure elements, which are now seen as
gas clouds responsible for weak Ly-$\alpha$ absorption lines situated
far from galaxies
($\approx$ 5 -- 6$h^{-1}$Mpc, Morris et al. 1993; Stocke et al.
1995; Shull et al. 1996) can also be considered as an evidence in
favor of large scale bias.

Simulations show also existence of
bias in the distribution of DM and 'galaxies' identified with the
highest peaks in the initial density distribution (see, e.g., Eke et al.
1996). Such bias is similar, in some important respects, to that generated
by reheating. It operates on really large scales (Bower et al. 1993) and
can be essential for the SLSS formation as well. It results in
structure composed of filaments and sheets in the distribution of
'galaxies' (Doroshkevich, Fong ~\&~ Makarova 1998). Other mechanisms
of bias formation discussed recently (Coles 1993; Sahni \& Coles 1995;
Tegmark \& Peebles 1998), operate on significantly smaller scales.

Simulations allow us to test the joint action of various factors on
the structure parameters and, therefore, they now seem to offer more
perspective way for detailed investigations of the structure
formation and evolution. They need, however, to be essentially
improved in order to discriminate the spatial distribution and other
parameters of 'galaxies' and dark matter. The methods used for the
measurement and description of simulated and observed structure
should be also improved as now they provide us with limited information
about structure properties. These restrictions become especially
important at higher redshifts, where the simulated density contrast is
small. Large dispersions of measured values make also any comparison
of simulated and theoretical structure parameters more
difficult. Nonetheless, such approach seems to be perspective because
further progress in these directions can be reached.

\subsection*{Acknowledgments}
We are grateful to our referee, Peter Coles, for very useful comments 
and criticism. 
This paper was supported in part by Denmark's Grundforskningsfond
through its support for an establishment of Theoretical Astrophysics
Center and Polish State Committee for Scientific
Research grant Nr. 2-P03D-022-10. AGD also wishes to acknowledge support
from the Center for Cosmo-Particle Physics "Cosmion" in the framework
of the project "Cosmoparticle Physics".

\medskip
\medskip
\centerline{\bf Appendix I}
\medskip
\centerline{\bf Correlation functions for initial perturbations.}
\medskip
To investigate mutual interaction of  perturbations
and to reveal their influence on the nonlinear processes of
structure formation we can use the conventional distribution
functions and conventional mean values.
In this appendix we introduce a few correlation and structure functions
which describe the relative spatial distribution of important parameters
of perturbations.

\begin{figure}
\centering
\epsfxsize=8 cm
\epsfbox{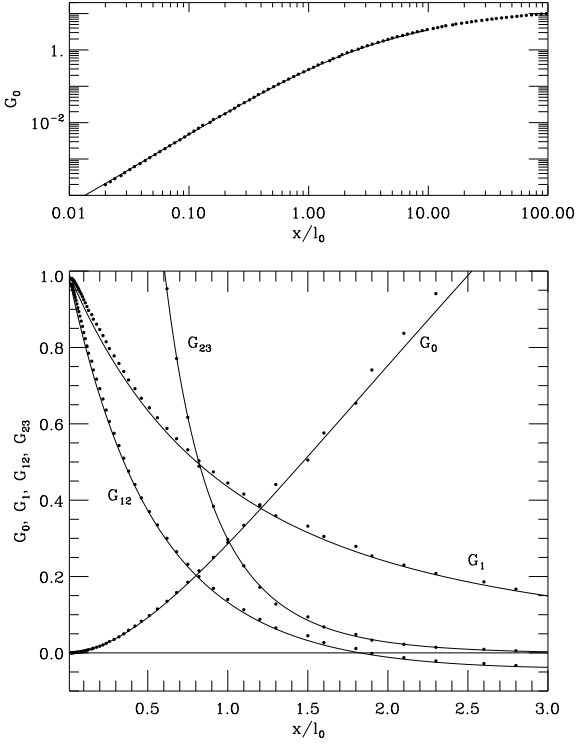}
\vspace{0.75cm}
\caption{Structure functions $G_0$ (top panel) and $G_0$,~$G_1$,~
$G_{12}$ ~\&~$-G_{23}$ vs. Lagrangian coordinate $q/l_0$ for the SCDM
power spectrum.
}
\end{figure}

We begin with the structure function of gravitational potential
perturbations which characterizes correlation of the gravitational
potential in two points ${\bf q_{1}}$ and ${\bf q_{2}}$. As the power
spectrum is a function of the absolute value of  wave number $|k|$ only
this structure function depends on $q_{12}=|{\vec q}_{1}-{\vec q}_{2}|$ 
and for the perturbations of gravitational potential we have
$$3{<\Delta\phi_{12}\Delta\phi_{34}>\over l_0^2\sigma_s^2} =
G_0(x_{14})-G_0(x_{13})+G_0(x_{23})-G_0(x_{24}),$$
$$\Delta\phi_{12} = \phi({\bf q}_1)-\phi({\bf q}_2),\quad
\Delta\phi_{34} = \phi({\bf q}_3)-\phi({\bf q}_4).\eqno(I.1)$$
Differentiation of this structure function gives other structure functions:
$${3<\Delta\phi(x_{12}) S_i(x_3)>\over l_0\sigma_{s}^2} =
(x_1-x_3)_iG_1(x_{13})-(x_2-x_3)_iG_1(x_{23}),$$
$${3<S_i(x_1)S_j(x_2)>\over \sigma_{s}^2}=\Delta_{ij}G_1(q_{12})+
{x_ix_j\over x^2}G_{12}(x))|_{x=x_{12}},$$
$${3l_0<S_k(x_1)d_{ij}(x_2)>\over\sigma_{s}^2} =\eqno(I.2)$$
$$(x_k\Delta_{ij}+x_i\Delta_{kj}+x_j\Delta_{ik})G_2(x)+
{x_ix_jx_k\over x^2}G_{23}(x))|_{x=x_{21}},$$
where the projection operator
$\Delta_{ij}=\delta_{ij}-x_ix_j/x^2$ is used and $x=q/l_0$. The scale
$l_0$ was introduced by (2.5).

In the general case these functions can be expressed through the
power spectrum, $p(k)$, and spherical Bessel functions and we
obtain
$$G_0(q) = {3\over 2\pi^2 l_0^2\sigma_{s}^2}\int_0^\infty
{p(k)\over k^2}\left(1-{\sin y\over y}\right)dk,\eqno(I.3)$$
$$G_1(q) = {3\over (2\pi)^{3/2}\sigma_{s}^{2}}
\int_0^\infty p(k)y^{-3/2}J_{3/2}(y)dk = G_0'/x,$$
$$G_2(q) = -{3l_0^2\over (2\pi)^{3/2}\sigma_{s}^2}
\int_0^\infty k^2p(k)y^{-5/2}J_{5/2}(y) dk= G_1'/x,$$
$$G_{3}(q) = {3l_0^4\over (2\pi)^{3/2}\sigma_{s}^2}
\int_0^\infty k^4p(k)y^{-7/2}J_{7/2}(y) dk= G_2'/x,$$
where $y=kq,\quad  x=q/l_0$.
$$G_{12}(x)=G_1(x)+x^2 G_2(x)=G_0^{''} = (xG_1)^{'},$$
$$G_{23}(x) = 3G_2(x)+x^2G_3(x) = G_{12}^{'}/x,\eqno(I.4)$$
$$G_{234}(x) = 3G_2(x)+6x^2G_3(x)+x^4G_4(x) = G_{12}^{''}.$$
$$G_1(0) = G_{12}(0) = 1,\quad G_{23}(0) = G_{234}(0) =
-{l_0^2\over l_c^2},\eqno(I.5)$$
and the scale $l_c$ was introduced by (2.5).

For realistic power spectra these functions cannot be expressed
analytically and they have to be calculated numerically. For the
Harrison - Zel'dovich primordial power spectrum and for the
CDM transfer function (BBKS), for 
$x=q/l_0\geq 6l_c^2/l_0^2$,~~ $6l_c^2/l_0^2\ll 1$ these functions can
 be approximated by 
$$G_0(x)\approx {1\over 2a_0}\ln(1+x+a_0x^2)$$
$$ -{1\over 2a_0\sqrt{4a_0-1}}tg^{-1}
\left({x\sqrt{4a_0-1}\over 2+x}\right),$$
$$G_1(x)\approx [1+x+a_0x^2]^{-1},$$
$$G_{12}(x)\approx G_1^2(x)[1-a_0 x^2],\eqno(I.6)$$
$$G_{23}(x)\approx -2G_1^3(x)(1+3a_0x^2-a_0^2x^3)/x,$$
$$G_{234}(x)\approx 6G_1^4(x)(1+1.3a_0x^2+4a_0^2x^3-a_0^3x^4),$$
$$a_0 = 5~(l_0/L_0)^2\approx 0.3, ~~ L_0\approx 4.1~l_0,$$
where the typical scale $L_0$ is defined as
$$L_0^2 = 3 \pi^2\int_0^\infty p(k)\left(1-{\sin{kL_0}
\over kL_0}\right)k^{-2}dk\Big/\int_0^\infty p(k) dk.$$
The functions (I.3) together with fits (I.6) are plotted in Fig. 7.
The more cumbersome expression
$$G_1(x)\approx (1+q_0)[1+\sqrt{q_0^2+x^2}+a_0x^2]^{-1},\eqno(I.7)$$
$$G_{12}\approx G_1^2(1+q_0)^{-1}[1-a_0x^2+q_0^2/\sqrt{q_0^2+x^2}]$$
$$q_0 = 6l_c^2/l_0^2[1+\sqrt{1+12 l_c^2/l_0^2}~]^{-1}\ll 1$$
can be used for all x. In this case the functions $G_{2}(x)$,
$G_{23}(x)$ and $G_{234}(x)$ can be found through the relations
(I.4).

\medskip
\medskip
\centerline{\bf Appendix II}
\medskip
\centerline{\bf Distribution function of pancakes.}
\medskip

We begin with the distribution function for the difference
of displacements at points with coordinates $\bf q_1$ and $\bf q_2$.
The normalization and notation introduced in Appendix I
are used. The dimensionless 'time' $\tau$ is introduced by (2.9).

Let us consider the deformation of spherical cloud with a diameter
$q$. For the case $q\gg q_0$ the general deformation of cloud can be
characterized by the 2D random scalar function $\Theta(\theta,\phi) =
[{\bf S}({\bf q}/2)-{\bf S}({\bf -q}/2)]\cdot {\bf q}/q^2$ instead of
the deformation tensor $d_{ik}$. Expansion of this function into
spherical harmonics allows one to obtain an approximate description of
deformation of the cloud with a reasonable accuracy. Now we will take
into account the spherical and quadrupole deformations only. These modes
describe successfully the transformation of the cloud into filaments
and sheets while higher order deformations relate mainly to the cloud
disruption.
In this case we can consider the deformation of the cloud in three
principal directions, namely, $x, y, \& z$. The dispersion of displacement
differences in these directions are
$$\sigma_x^2=<[S_x(q/2)-S_x(-q/2)]^2>= 2[1-G_{12}(q)],\eqno(II.1)$$
$$\sigma_y^2 = \sigma_z^2 = \sigma_x^2$$
and their correlations are described by the coefficient $r_{xy}$:
$$\sigma_x^2 r_{xy}=<[S_x(q/2)-S_x(-q/2)][S_y(q/2)-S_y(-q/2)]>$$
$$r_{xy}=-{q^2\over 2}{G_2(q/\sqrt{2})\over 1-G_{12}(q)}
\approx {0.25\sqrt{2}\over 1+q/2}\left({1+q\over
1+q/\sqrt{2}}\right)^2,\eqno(II.2)$$
for $q_0\ll q$.
Taking into account the approximate character of our consideration
and because of $r_{xy} = r_{xz} =r_{zy}\ll$1 we will neglect these
correlations and will consider the deformation along three principal
axes as independent. Under this assumption we have for the
distribution of maximal displacement difference
$\Delta S_x\geq \Delta S_y\geq \Delta S_z$ and for 
$\zeta=\Delta S_x/\sigma_x$
$${dW(\zeta)\over d\zeta} = {3\over 4\sqrt{2\pi}}e^{-\zeta^2/2}
\left[1+erf\left({\zeta\over\sqrt{2}}\right)\right]^2,~
\zeta\geq 0,\eqno(II.3)$$
$${dW(\zeta)\over d\zeta} = {3\over 4\sqrt{2\pi}}e^{-\zeta^2/2}
\left[1-erf\left({\zeta\over\sqrt{2}}\right)\right]^2,~
\zeta\leq 0.$$
These relations show that under the
above assumptions $\sim$ 1/8 of matter is compressed in all
directions and $\sim$ 1/8 expanded in all directions, whereas $\sim$3/8
of matter is compressed along $x \& y$ axes and $\sim$3/8 of matter
is compressed along $x$ axis and expanded along $y \& z$ axes.
This means that 7/8 of matter is compressed at least along one of the
axes. The influence of correlation does not distort the symmetry
between compression and expansion and, at $q\leq$ 1, increases
the probability to find
$\Delta S_x\geq \Delta S_y\geq\Delta S_z\geq$ 0 up to 0.21.

Let us now consider the matter compression along the direction
of maximal compression $\bf q = q_1-q_2$. In this case
the dispersions of displacement differences
$\Delta S_i=S_i(q_1)-S_i(q_2)$ are:
$$\sigma_{11}^2=<\Delta S_1\Delta S_1> = 2[1-G_{12}(q)],\eqno(II.4)$$
$$\sigma_{22}^2=<\Delta S_2\Delta S_2> =
\sigma_{33}^2=<\Delta S_3\Delta S_3> = 2[1-G_1(q)],$$
$$<\Delta S_1\Delta S_2> = <\Delta S_1\Delta S_3> =
<\Delta S_2\Delta S_3> =0.$$

For the distribution function (II.2) the probability to have
$\Delta S_1\geq q/\tau$ is
$$W_{cr}(q) = 1-{1\over 8}\left[1+erf\left({\mu(q)\over
\sqrt{2}\tau}\right)\right]^3,\eqno(II.5)$$
$$\mu(q)={q\over \sqrt{2[1-G_{12}(q)]}}.$$
This function characterizes the intersection of two particles
before the time moment $\tau$ or for the pancake separation $\geq q$
and, in fact, $(8/7)W_{cr}$ is the cumulative distribution function
for pancakes with masses $>q$, for a given $\tau$.

The standard technique (see, e.g., BBKS) can be used in order to
find the mean comoving linear density of pancakes with a given
$\Delta S_1=q/\tau$ along a random straight line. It is expressed
through the characteristics of  the derivations of function $Q$
introduced by (3.2), $Q_i=\partial Q/\partial q_i$, and for $q_0\ll$1
$$<Q_i>=0,\quad \sigma_{d1}^2 = <Q_1^2>\approx 6/q_0,\eqno(II.6)$$
$$\sigma_{d2}^2=<Q_2^2>=\sigma_{d3}^2=<Q_3^2>\approx 2/q_0$$
($q_0$ is introduced by (I.7)).
The required linear number density, $n(>q)$, is
$$l_0n(>q) = {3<\mu_r>\over 16\pi^2\sigma_{11}\sigma_{d1}\sigma_{d2}
\sigma_{d3}}\cdot\Phi\left({\mu(q)\over\sqrt{2}\tau}\right)\times
\eqno(II.7)$$
$$\int\sqrt{Q_1^2+Q_2^2+Q_3^2}\exp\left[-{Q_1^2\over 2\sigma_{d1}^2}-
{Q_2^2\over 2\sigma_{d2}^2}-{Q_3^2\over 2\sigma_{d3}^2}\right]
dQ_1dQ_2dQ_3$$
where $\mu_r$ characterizes the random angle of the intersection.
For $l_c/l_0\ll 1$, and $q\gg q_0$, we obtain (3.10).

\medskip
\medskip
\centerline{\bf Appendix III}
\medskip
\centerline{\bf Interactions of pancakes.}
\medskip

In order to investigate the correlations of the pancake properties
we need to consider the intersection of the points $q_1$ and $q_2$,
~$q_{12}=q_1-q_2$, at the moment $\tau_1$, and points $q_3$ and $q_4$,
~$q_{34}=q_3-q_4$, at the moments $\tau_2$, respectively. We consider
the simplest case when the vectors ${\bf q}_{12}$ and ${\bf q}_{34}$
are situated along the same line.

It is convenient to characterize the pancakes by their sizes, $D_1$
and $D_2$, and by the separation of their central points, $D_{sep}$.
In this case we have
$$q_{12}=D_1,\quad q_{34}=D_2,$$
$$q_{31}=D_{sep}-0.5(D_1-D_2),\quad q_{41}=D_{sep}-0.5(D_1+D_2),$$
$$q_{32}=D_{sep}+0.5(D_1+D_2),\quad q_{42}=D_{sep}+0.5(D_1-D_2).$$
The dispersions and the correlation of  displacements
$\Delta S(D_1)$ and $\Delta S(D_2)$ can be written as
$$\sigma_1^2(D_1) = \sigma_{11}(D_1),\quad
\sigma_2^2(D_2) =  \sigma_{11}(D_2),\eqno(III.1)$$
$$\sigma_1\sigma_2~r_s = G_{12}(q_{31})-G_{12}(q_{32})+
G_{12}(q_{42})-G_{12}(q_{41}),$$
and $r_s$ is a symmetric function of $D_{sep}$.
For $D_{sep}\rightarrow \infty\quad r_s\rightarrow 0$, and
for $D_{sep}=0$
$$\sigma_1\sigma_2r_s = 2\left[G_{12}\left({D_1-D_2\over 2}\right)-
G_{12}\left({D_1+D_2\over 2}\right)\right],   \eqno(III.2)$$
$$r_s(D_1,D_2,0)\approx -\mu(D_1)\mu(D_2)G_{23}(D_1/2),\quad D_2\ll 1.$$
For $D_{sep}=0,~~D_1=D_2, ~~r_s=1$.

For a given $\Delta S(D_1)=D_1/\tau_1$ the conditional mean values
$<\Delta S_c(D_2,D_{sep})>$ and dispersions $\sigma_c(D_2,D_{sep})$
are described by the standard relations
$$<\Delta S_c(D_2,D_{sep})> = \sigma_2 r_s\mu_1/\tau_1,~
\sigma_c(D_2,D_{sep}) = \sigma_2\sqrt{1-r_s^2},$$
and the conditional probability of
the pancake formation with $\Delta S(D_2,D_{sep})\geq D_2/\tau_2$
under the condition $\Delta S(D_1)= D_1/\tau_1$ is
$$W_c(>D_2, \tau_2;D_1,\tau_1,D_{sep})  =
0.5 erfc\left({g_2\over \sqrt{2}}\right).\eqno(III.3)$$
$$g_2 = \left({\mu(D_2)\over \tau_2}-r_s{\mu(D_1)\over
\tau_1}\right)(1-r_s^2)^{-1/2}.$$
These relations allow us to describe the formation of pancakes
with $D_2\leq D_1$ at the moment $\tau_2\leq \tau_1$ and, in
particular, to find the function $<m_{cold}(D_{sep})>$ and
$N_m(D_{sep})$. For $\tau_1 = \tau_2$ the same relations 
describe the accumulation of the smaller pancake $(D_2)$ by 
the larger one $(D_1)$.

More complicated relations describe the pancake coagulation. As it
was noted in Sec. 3.4 this process is characterized by the function
$$\psi(D_1,D_2,D_{sep}) = {\phi(q_{cent}+D_1/2)-
\phi(q_{cent}-D_1/2)\over D_1} - $$
$${\phi(D_{sep}+q_{cent}+D_2/2)-
\phi(D_{sep}+q_{cent}-D_2/2)\over D_2},$$
and the condition of coagulation is
$$\psi(D_1,D_2,D_{sep}) = D_{sep}/\tau.\eqno(III.4)$$
To obtain the statistical description of this process the conditional
mean value $<\psi^c(D_1,D_2,D_{sep})>$ and dispersion $\sigma_\psi^c$
need to be found. Using the expressions given in Appendix I and
notation introduced above we obtain:
$${1\over 2}\sigma_\psi^2 = {G_0(D_1)\over D_1^2}+{G_0(D_2)\over D_2^2}+$$
$${G_0(q_{31})-G_0(q_{41})-G_0(q_{32})+G_0(q_{42})\over D_1D_2}.
\eqno(III.5)$$
The cross correlation of functions $\psi$ and $\Delta S(q_{12}),~
\Delta S(q_{34})$ is described as following:
$${<\psi \Delta S(D_1)>\over \sigma_\psi\sigma_1} = \mu(D_1)r_\psi,~
{<\psi \Delta S(D_2)>\over \sigma_\psi\sigma_2} = \mu(D_2)r_\psi,$$
$$r_\psi = {q_{31}G_1(q_{31})-q_{41}G_1(q_{41})-
q_{32}G_1(q_{32})+q_{42}G_1(q_{42})\over D_1D_2\sigma_\psi}.$$
For the conditional mean value and dispersion of $\psi$ we have
$${<\psi^c>\over \sigma_\psi} = r_\psi\left({(\mu(D_1)-\mu(D_2))^2\over
1-r_s^2} +{2\mu(D_1)\mu(D_2)\over 1+r_s}\right),$$
$$\left({\sigma_\psi^c\over \sigma_\psi}\right)^2 = 1-r_\psi^2
\left({(\mu(D_1)-\mu(D_2))^2\over 1-r_s^2} +
{2\mu(D_1)\mu(D_2)\over 1+r_s}\right).$$
Finally for the required function $\chi(D_1,D_2,D_{sep})$ we obtain
$$\chi = [D_{sep} - <\psi^c>]/\sigma_\psi^c.\eqno(III.6)$$

\medskip
\medskip
\centerline{\bf Appendix IV}
\medskip
\centerline{\bf The dynamical characteristics of pancakes.}
\medskip
A correlation of pancake velocity described by (8.4) in points
with different transversal coordinates is conventionally characterized
by the correlation coefficient
$$r_v={<[\phi({\bf q}_1)-\phi({\bf q}_2)][\phi({\bf q}_3)-
\phi({\bf q}_4)]>\over\sqrt{<[\phi({\bf q}_1)-\phi({\bf q}_2)]^2>
<[\phi({\bf q}_3)-\phi({\bf q}_4)]^2>}}=$$
$${G_0(q_{14})-G_0(q_{13})+G_0(q_{23})-G_0(q_{24})\over
2\sqrt{G_0(q_{12})G_0(q_{34})}}.\eqno(IV.1)$$
For small $q_{12}, \& q_{34}$ it transforms into (8.8).

The important characteristic of the pancake is the velocity
profile around the points $q_1 \& q_2$ under the condition
$\Delta S(q_{12})=q_{12}/\tau$. It is described by the mean
conditional profile of displacement
$$<S(q_3)> = {q_1-q_2\over \tau}{G_{12}(q_{13}) - G_{12}(q_{23})
\over 2[1-G_{12}(q_{12})]},\eqno(IV.2)$$
and the conditional dispersion
$$\sigma_s^2(q_3) = 1-{[G_{12}(q_{13}) - G_{12}(q_{23})]^2\over
2[1-G_{12}(q_{12})]}.\eqno(IV.3)$$
Using the expression (8.9) the velocity dispersion within pancakes
can be found as follows:
$${(1+z)^2\over H^2(z)}\sigma_{v}^2 =
{q_{12}^2\over 12}[(1-\beta(z))^2+2\beta(z)f_1(q_{12})- $$
$$\beta^2(z)f_2(q_{12})+\tau^2\beta^2(z)f_3(q_{12})],\eqno(IV.4)$$
$$f_1(q_{12})=1-{12\over q_{12}^2}{G_0(q_{12})-0.5q_{12}^2G_1(q_{12})
\over 1-G_{12}(q_{12})},$$
$$f_2(q_{12})=1-{3J_{12}\over [1-G_{12}(q_{12})]^2}, $$
$$f_3(q_{12})=1-{J_{12}\over 2 [1-G_{12}(q_{12})]}-
{2G_0(q_{12})\over q_{12}^2},$$
$$J_{12}=q_{12}^{-1}\int_{q_2}^{q_1} dq_3[G_{12}(q_1-q_3)-
G_{12}(q_2-q_3)]^2.$$

\end{document}